\documentclass[12pt,a4paper]{article}
\usepackage[portuguese]{babel}

\usepackage[a4paper, top=3cm,bottom=3cm,left=3cm,right=3cm,marginparwidth=1.75cm]{geometry}
\usepackage{amsmath,amsfonts,amsthm}
\usepackage{natbib}
\usepackage{graphicx}
\usepackage[colorlinks=true, allcolors=blue]{hyperref}
\usepackage{xurl}
\usepackage{enumitem} 
\usepackage{authblk}
\usepackage{multirow} 

\title{\Large \bf Mensuração da Transferência de Riqueza em Planos de Contribuição Definida com a Marcação de Ativos na Curva}

\author[1]{Eduardo Fraga L. de Melo}
\author[2]{Rodrigo S. Targino}

\affil[1]{\small Instituto de Matemática e Estatística - IME, UERJ, Rio de Janeiro}
\affil[2]{\small Escola de Matemática Aplicada - EMAp, FGV, Rio de Janeiro}

\renewcommand\and{ e }

\date{}  

\begin{document}

\maketitle

\begin{abstract}

\noindent O método de mensuração de ativos financeiros em planos de previdência na modalidade de contribuição definida (CD, ou contribuição variável - CV, na fase de acumulação) tem implicações significativas se haverá transferência de riqueza entre os participantes. Em Dez/2024 foi publicada norma para as Entidades Fechadas de Previdência Complementar possibilitando o uso da marcação na curva de títulos públicos federais nos planos CD e CV na fase de acumulação. Este artigo quantifica o impacto financeiro nos participantes da adoção da marcação na curva (HTM – {\it Hold to Maturity}) nestes planos, utilizando dados reais da estrutura a termo da taxa de juros de cupom de IPCA para avaliar as transferências de riqueza resultantes dessa adoção. A análise evidencia como a marcação na curva gera assimetrias nos resultados financeiros, beneficiando alguns participantes em detrimento de outros. As transferências de riqueza ocorrem tanto em qualquer retirada de recursos quanto também na entrada (contribuições), inclusive realocações da carteira que impliquem venda ou compra de títulos. O uso do HTM de forma parcial ou a tentativa de imunização de saídas não eliminam por completo transferências de riqueza. Os resultados reforçam que, para fins de cotização, o uso da marcação a mercado (MTM - {\it Mark to Market}) de ativos em planos CD (e CV na fase de diferimento) evita transferências de riqueza e, por consequência, prejuízos financeiros aos seus participantes.

\ \

\noindent \textbf{Palavras-chaves}: planos CD; marcação de ativos na curva (HTM); marcação de ativos a mercado (MTM); transferência de riqueza.

\ \

\noindent \textbf{Classificação JEL}: C00; C60.




\end{abstract}

\newpage

\section{Introdução}

A mensuração de ativos financeiros desempenha um papel fundamental na alocação eficiente de recursos e na transparência do sistema financeiro. No contexto dos planos de previdência complementar na modalidade de Contribuição Definida (CD), a metodologia utilizada para mensurar os ativos, em específico os de renda fixa, e calcular o valor das cotas influencia diretamente se haverá ou não transferência de riqueza entre os participantes. A escolha entre a marcação a mercado (MTM – {\it Mark to Market}) e a marcação na curva (HTM – {\it Hold to Maturity}) é um dos pontos centrais com impactos significativos sobre a equidade nesses planos.

A marcação a mercado (MTM) reflete o preço corrente dos ativos financeiros, ajustando constantemente os valores dos investimentos de acordo com as variações das condições de mercado (\cite{Bodie2018}). Essa abordagem costuma ser utilizada para garantir transparência e evitar distorções na valoração dos ativos. Em contrapartida, a marcação na curva (HTM) estabelece que os ativos sejam precificados com base no fluxo de caixa do título utilizando taxas de juros contratadas no momento da aquisição. Conforme menciona \cite{Fabozzi2013}, embora essa metodologia reduza a volatilidade aparente dos investimentos, ela pode ocultar perdas ou ganhos que seriam evidentes sob uma abordagem de MTM.

Em planos CD, onde os participantes arcam com o risco dos investimentos (\cite{BlakeCairns2001}), a valoração adequada dos ativos é essencial para assegurar uma distribuição correta dos recursos. A adoção da marcação na curva pode criar distorções ao permitir que participantes que adquiram (aportes/contribuições) ou retiram suas cotas (seja por morte, invalidez, aposentadoria, perda do vínculo ou outro) em momentos distintos sejam impactados de forma desigual, gerando transferências de riqueza dentro do plano. Isso ocorre porque os valores das cotas deixam de refletir a realidade econômica dos ativos subjacentes, podendo beneficiar participantes, por exemplo, que retiram ou realocam recursos, gerando a necessidade de venda de títulos, em detrimento daqueles que permanecem no plano ou o contrário, prejudicando participantes que retiram recursos em benefício daqueles que permanecem no plano. Neste artigo, estão incluídos, nos conceitos de ``entrada'' e ``saída'' de recursos, as realocações de investimento que impliquem na compra ou venda de títulos públicos federais, ou seja, eventuais trocas de ``perfis de investimento'' feitas por participantes também implicam em transferências de riqueza caso sejam feitas negociações destes títulos em plano que adote HTM. Além dessas razões, realocações podem ser feitas em função de ajustes na carteira devido à busca pela imunização de expectativa, ou outra medida de risco, de retiradas futuras.

No contexto em que os participantes carregam o risco de retorno dos ativos nos planos CD, \cite{BlakeCairns2001} estimam valores em risco ($VaR$) desses retornos na fase de acumulação. De acordo com os autores, o fator de risco desses planos são os retornos de ativos, que levam em consideração, inclusive, estratégias de alocação (estáticas e com formas dinâmicas simples, como estilo de vida, limite e seguro de portfólio de proporção constante). A estimação de $VaR$ na fase de acumulação de planos CD implica que ativos são mensurados a mercado. Além disso, os autores citam que, nestes planos, ``o participante assume o risco de preço dos ativos (o risco de perdas no valor do seu saldo devido a quedas nos valores dos ativos, inclusive no período imediatamente anterior à aposentadoria)''. É papel do gestor do plano gerir/mitigar tal problema em momentos próximos à aposentadoria sem trazer prejuízos financeiros aos demais participantes por transferências de riqueza.

A regulamentação em vigor (Resolução CNPC 61/2024, de 11/12/2024) passou a permitir que planos CD e CV, na fase de diferimento, de Entidades Fechadas de Previdência Complementar (EFPC) utilizem a marcação na curva (HTM) de títulos públicos federais. Diante desse contexto, este artigo busca quantificar as transferências de riqueza decorrentes dessa adoção, utilizando dados reais da estrutura a termo da taxa de juros (ETTJ) de cupom de IPCA. O uso destas taxas aproxima o estudo da realidade da variação, entre os anos, dos preços dos ativos. Não são taxas de juros hipotéticas nem simuladas. Portanto, os resultados numéricos de transferência de riqueza seriam reais para o período de análise para os casos desenvolvidos.

A análise realizada pretende evidenciar as assimetrias financeiras geradas por essa prática. Como exposto no artigo, as transferências de riqueza ocorrem tanto em qualquer retirada de recursos quanto também na entrada. Logo, incluem retiradas por morte, invalidez, aposentadoria, resgates, portabilidades, aportes, contribuições e realocações de investimentos que impliquem compras ou vendas de títulos públicos federais.

Conforme consta de \cite{Previc2024}, no relatório mais recente disponível no sítio eletrônico em consulta feita em 15/03/2025, os planos CD possuem R\$ 184 bilhões em ativos ao fim de 2023. Sobre os planos CV, há informação de que 64\% do total das provisões matemáticas está em benefícios a conceder (fase de acumulação). Aplicando este percentual no total de ativos em planos CV (R\$ 356 bilhões), tem-se o valor estimado de R\$ 227 bilhões ao fim de 2023. Dessa forma, o total de ativos em planos CD e CV, na fase de acumulação, seria de R\$ 411 bilhões. Se considerarmos a parcela investida em renda fixa (aprox. 80\%, conforme relatório), e dentro desta, a investida em títulos públicos federais, aprox. 90\%, conforme relatório (no documento não foi encontrada esta parcela para planos CD e CV, separadamente para a fase de diferimento), ter-se-iam 297 bilhões de reais, em valores aproximados, expostos à marcação na curva.

O artigo está organizado da seguinte forma: na seção 2, as normas de mensuração de ativos em planos CD serão reportadas, assim como os critérios de mensuração HTM e MTM serão mais detalhados com exemplos simplificados. Na seção 3, a metodologia feita com dados reais de estrutura a termo de taxa de juros de cupom de IPCA entre 2005 e 2024 será apresentada com a implementação de diversos cenários para planos CD. Os resultados deste estudo serão mostrados na seção 4. Por fim, considerações finais são feitas.

\section{Planos CD e mensuração de ativos}

Esta seção é iniciada com breve apresentação do plano de contribuição definida (CD), cujas características são relevantes para a compreensão dos efeitos da forma de mensuração de ativos. Posteriormente, apresentamos as mudanças normativas nos últimos anos que terminam com a alteração que possibilitou que estes planos utilizassem a marcação na curva. O plano CD possui estrutura equivalente ao CV na fase de contribuição.

Uma vez caracterizado o plano CD e apresentadas as mudanças ocorridas nos critérios para mensuração de títulos públicos federais, faremos uma ilustração numérica das metodologias para identificar e mensurar a transferência de riqueza ao se usar a marcação de ativos na curva (HTM).

\subsection{Planos CD}

A aposentadoria em um plano CD é baseada no valor do montante acumulado durante a fase de diferimento (ou seja, durante a vida profissional do participante) das contribuições que vão para o plano e dos retornos de investimento sobre elas. Essas contribuições podem ser pagas apenas pelo participante, apenas pelo empregador ou por uma combinação dos dois.

Nestes planos, o participante realiza aportes mensais determinados por um percentual do salário. O valor do benefício futuro, seja como renda mensal por meio de retiradas programadas ou o saldo acumulado, permanece incerto até o momento da aposentadoria, pois dependerá tanto do total contribuído ao longo do tempo quanto do desempenho dos investimentos. Para outros detalhes sobre esses planos, ver \cite{SCHAUS2017}.

Dessa forma, sobre estas características apenas, o plano CD se aproxima, na essência, de um fundo de investimento: o saldo é individualizado, cotizado, com lastro em ativos e seu valor varia ao longo do tempo conforme os rendimentos destes ativos e os novos aportes. Os fundos de pensão realizam a cotização, ou seja, calculam periodicamente o valor da cota, permitindo aos participantes acompanhar a evolução da rentabilidade dos recursos do plano.

No que se refere às finanças em fundos de pensão e, especificamente, em planos CD, referenciamos o cap. 8 de \cite{Blake2006}, onde o autor sugere e compara performance de planos CD por meio dos retornos observados das carteiras, ou seja, por meio do efetivo uso da marcação a mercado ainda que sejam planos com ativos de alta volatilidade (no caso do capítulo do livro, ações, que costumam ser bem mais voláteis que títulos de renda fixa).

A diferença entre planos CD e fundos de investimento se dá nas possibilidades de retirada de recursos. Nos planos, os recursos podem sair em caso de morte, invalidez, aposentadoria ou portabilidade/resgate (em caso de mudança de emprego ou perda do vínculo com o empregador e referente ao montante acumulado de contribuições do empregado). Sobre a portabilidade ou o resgate, o montante acumulado a que se refere eventuais contribuições do empregador deve observar condições do regulamento. Apesar de, em tese, não haver tanta flexibilidade na saída de recursos, se observa que saídas sempre existirão em uma massa de participantes de um plano CD em um horizonte de tempo. Além disso, há também as realocações de investimentos.

De acordo com \cite{Bri2006}, por definição, um plano CD é sempre totalmente coberto (afinal ativo e passivo caminham perfeitamente juntos) e o empregador normalmente não tem nenhuma obrigação financeira além de fazer pagamentos periódicos no plano. Os empregadores que oferecem planos CD escapam dos riscos financeiros e de longevidade associados aos planos de Benefício Definido (BD). Os autores desse estudo afirmam que {\it employers offering DC plans escape financial and longevity risks associated with DB payments, but then are denied access to the opaque accounting of pension assets and liabilities in DB plans common in many countries, including the US, Canada, and until recently the UK}. Antes das reformas que impuseram medidas para mensuração mais baseada em valores de mercado para planos BD, os autores citam que {\it pension accounting seems to have, at times, allowed some firms to obscure the costs of their pension plans and even the actual volatility and level of net income}.

\subsection{Normas sobre mensuração de ativos de planos CD}

A Resolução CNPC 37/2020 e, posteriormente, a Resolução CNPC 43/2021\footnote{https://www.in.gov.br/en/web/dou/-/resolucao-cnpc-n-43-de-6-de-agosto-de-2021-341365068} estabeleciam que a mensuração de títulos públicos federais na curva (ou seja, a categoria HTM - mantidos até o vencimento) poderia ser utilizada nos planos de Benefício Definido, caso se demonstrasse que o plano de benefícios possui intenção e capacidade financeira de mantê-los até o vencimento, e que o prazo entre a data de aquisição e a data de vencimento dos títulos fosse igual ou superior a cinco anos. Para planos CD e CV, na fase de acumulação, deveria ser utilizada a marcação a mercado (MTM). A redação do dispositivo em questão era:\\

\noindent {\it § 2º A entidade pode registrar os títulos públicos federais na categoria "títulos mantidos até o vencimento" em planos de benefícios na modalidade de benefício definido, quando o prazo entre a data de aquisição e a data de vencimento dos títulos for igual ou superior a cinco anos e desde que haja capacidade financeira e intenção em mantê-los na carteira até o vencimento.}\\

Entretanto, a Resolução CNPC 61/2024\footnote{https://www.in.gov.br/en/web/dou/-/resolucao-cnpc-n-61-de-11-de-dezembro-de-2024-602263594} alterou essa diretriz, permitindo que planos CD e CV, na fase de acumulação, pudessem utilizar a marcação na curva (HTM) de títulos públicos federais, caso se demonstre que o plano de benefícios possui intenção e capacidade financeira de mantê-los até o vencimento, e o prazo entre a data de aquisição e a data de vencimento dos títulos seja igual ou superior a cinco anos. A redação do dispositivo em questão passou a ser:

\ \

\noindent {\it § 2º A entidade pode registrar os títulos públicos federais na categoria "títulos mantidos até o vencimento" se atendidas as seguintes condições:\\
I - demonstrar que o plano de benefícios possui intenção e capacidade financeira de mantê-los até o vencimento; e\\
II - o prazo entre a data de aquisição e a data de vencimento dos títulos for igual ou superior a cinco anos.}\\

Este artigo quantificará os impactos financeiros desta medida para os participantes de planos CD ou CV, na fase de acumulação.


\subsection{HTM: transferência de riqueza na entrada de recursos}

Para ilustrar em números a transferência de riqueza na entrada de recursos no plano, faremos três exemplos. No primeiro, há apenas um único tipo de título público. No segundo, há dois diferentes títulos públicos. O terceiro é igual ao segundo, mas com marcação na curva (HTM) feita apenas em parte dos títulos (por ex. $30\%$). Neste último exemplo, a outra parcela é marcada a mercado (MTM).
Esses exemplos são importantes para ilustrar como é feita a mensuração da transferência de riqueza. Em todos os exemplos:

\begin{itemize}
    \item Ana entra no plano com \$1 de contribuição no tempo $t=1$ e \$1 de contribuição no tempo $t=2$.
    \item Marcos apenas entra no plano no tempo $t=2$ com \$1 de contribuição.
    \item O valor da cota em $t=1$ é 1, tanto para um plano que adote MTM quanto para um plano que adote HTM.
    \item O valor do TítuloA (único título no exemplo 1) em $t=1$ é \$1.  \\
    No tempo $t=2$, o valor a mercado deste título passa a ser \$0,90. \\ 
    No tempo $t=2$, o valor na curva deste título passa a ser \$1,01.
    \item O valor do TítuloB (usado apenas nos exemplos 2 e 3) em $t=1$ é \$0,90.  \\
    No tempo $t=2$, o valor a mercado deste título passa a ser \$0,85. \\ 
    No tempo $t=2$, o valor na curva deste título passa a ser \$0,91.
    \item No exemplo 2, a alocação das contribuições será sempre 50\% no TítuloA e 50\% no TítuloB para facilitar, embora a alocação possa ser qualquer uma.
\end{itemize}

\subsubsection{Exemplo 1}

\noindent {\bf Situação MTM}

\begin{enumerate}[label=\alph*.]
    \item No tempo $t=1$, Ana contribuiu com \$1, comprou 1 cota, que equivale a 1 título, que custa \$1.
    \item Como o plano possui um único título, no tempo $t=2$, o valor da cota deste plano passa a ser $0,90 = 1 \times \frac{1 \times \$0,90}{1 \times \$1,00}$ (cotização considerando valores a mercado).
    \item No tempo $t=2$, Ana contribui com mais \$1, que compra $1/0,90$ cotas ($= 1,11$ cotas) que equivalem a $1/0,90$ títulos ($= 1,11$ títulos que o plano comprou no mercado).\\
    Ana passa a ter $2,11$ cotas ($= 1 + 1,11$ cotas) que equivalem a $2,11$ títulos ($= 1 + 1,11$ títulos).
    \item No tempo $t=2$, Marcos contribui com \$1, que compra $1/0,90$ cotas ($= 1,11$ cotas) que equivalem a $1/0,90$ títulos ($= 1,11$ títulos que o plano comprou no mercado).
    \item O plano possui valor total de $\$2,90$ ($2,11 + 1,11 = 3,22$ cotas; a $0,90$ o valor da cota, ou seja, $3,22 \times 0,90 = 2,90$).
    \item Ana possui:\\
    $65,52\%$ das cotas do plano ($=2,11/(2,11+1,11)$) que equivalem a $65,52\%$ dos títulos ($=2,11/(2,11+1,11)$).\\
    Marcos possui:\\
    $34,48\%$ das cotas do plano ($=1,11/(2,11+1,11)$) que equivalem a $34,48\%$ dos títulos ($=1,11/(2,11+1,11)$).
\end{enumerate}

Não há diferença entre a razão de cotas e a razão de títulos para nenhum dos participantes.
Marcos, ao entrar no plano, não tornou essa diferença (razão de títulos $-$ razão de cotas) positiva nem negativa para Ana.

Vamos analisar no caso HTM agora.

\ \

\noindent {\bf Situação HTM}

\begin{enumerate}[label=\alph*.]
    \item No tempo $t=1$, Ana contribuiu com \$1, comprou 1 cota, que equivale a 1 título, que custa \$1.
    \item No tempo $t=2$, o valor da cota deste plano passa a ser $1,01$ (seguindo a curva do título comprado no tempo $t=1$).
    \item No tempo $t=2$, Ana contribui com mais \$1, que compra $1/1,01$ cotas ($= 0,99$ cotas) que equivalem a $1/0,90$ títulos ($= 1,11$ títulos), afinal o título foi adquirido no mercado no tempo $t=2$ ao preço de $\$0,90$.\\
    Ana passa a ter $1,99$ cotas ($=1+0,99$ cotas) que equivalem a $2,11$ títulos ($=1+1,11$ títulos).
    \item No tempo $t=2$, Marcos contribui com \$1, que compra $1/1,01$ cotas ($= 0,99$ cotas) que equivalem a $1/0,90$ títulos ($= 1,11$ títulos), afinal o título foi adquirido no mercado no tempo t=2 ao preço de $\$0,90$.
    \item O plano possui valor contábil total de $ \$3,01$ ($2,98$ cotas; a $1,01$ o valor da cota).
    \item Ana possui:\\
    $66,78\%$ das cotas do plano ($=1,99/(1,99+0,99)$) que equivalem a $65,52\%$ dos títulos ($=2,11/(2,11+1,11)$).\\
    Marcos possui:\\
    $33,32\%$ das cotas do plano ($=0,99/(1,99+0,99)$) que equivalem a $34,48\%$ dos títulos ($=1,11/(2,11+1,11)$).
\end{enumerate}

Há diferença entre a razão de cotas e a razão de títulos para ambos os participantes.
Marcos, ao entrar no plano, fez com que Ana tivesse um percentual de cotas no plano maior que seu percentual de títulos. Já Marcos está com um percentual de cotas no plano menor que seu percentual de títulos.

Ao entrar no plano, Marcos transferiu riqueza para Ana. 

O valor dessa transferência é igual a: $(33,32\% - 34,48\%) \times 2,90 = - \$0,037$. 
Em termos percentuais, equivale a $3,7\%$ de sua contribuição.\\

Ana recebeu, em transferência de riqueza, $(66,78\% - 65,52\%) \times 2,90 = \$0,037$ no tempo $t=2$. 
Em termos percentuais, equivale a 3,7\% de sua contribuição ou 1,26\% do seu saldo no plano no tempo $t=2$ (calculado em relação ao valor usando MTM).

\subsubsection{Exemplo 2}

Neste exemplo, a alocação das contribuições será sempre 50\% no TítuloA e 50\% no TítuloB.

\ \

\noindent {\bf Situação MTM}

\begin{enumerate}[label=\alph*.]
    \item No tempo $t=1$, Ana contribuiu com \$1. comprou 1 cota.\\
    $\$0,50$ foi destinado a comprar TítuloA ($50\% \times 1 / 1 = 0,500$ títulos) e\\
    $\$0,50$ foi destinado a comprar TítuloB ($50\% \times 1/ 0,90 = 0,556$ títulos).
    \item No tempo $t=2$, o valor da cota deste plano passa a ser $0,9222$ ($=1 \times \frac{0,500 \times 0,90 + 0,556 \times 0,85}{0,500 \times 1,00 + 0,556 \times 0,90}$).
    \item No tempo $t=2$, Ana contribui com mais $\$1$, que compra $1/0,9222$ novas cotas ($= 1,08434$ cotas) que equivalem a:\\
    $50\% \times 1/0,90$ TítuloA ($= 0,556$ títulos) e\\
    $50\% \times 1/0,85$ TítuloB ($= 0,588$ títulos).\\
    Ana passa a ter $2,08434$ cotas ($= 1 + 1,08434$ cotas) que equivalem a:\\
    $1,056$ TítuloA ($= 0,500 + 0,556$ títulos) e\\
    $1,144$ TítuloB ($= 0,556 + 0,588$ títulos).
    \item No tempo $t=2$, Marcos contribui com \$1, que compra $1/0,9222$ cotas ($= 1,08434$ cotas) que equivalem a:\\
    $50\% \times 1/0,90$ TítuloA ($= 0,556$ títulos) e\\
    $50\% \times 1/0,85$ TítuloB ($= 0,588$ títulos).
    \item Plano possui valor total de $\$2,92$ ($2,08434 + 1,08434 = 3,16867$ cotas, a $0,9222$ o valor da cota).
    \item Ana possui $65,78\%$ das cotas do plano ($=2,08434/(2,08434+1,08434)$). Seus recursos dentro do plano são responsáveis por:\\
    $65,52\%$ do TítuloA ($=1,05556/(1,05556+0,55556)$)\\
    $66,04\%$ do TítuloB ($=1,14379/(1,14379+0,58824)$)\\
    \item Marcos possui 34,22\% das cotas do plano ($=1,08434/(2,08434+1,08434)$). Seus recursos dentro do plano são responsáveis por:\\
    $34,48\%$ do TítuloA ($=0,55556/(1,05556+0,55556)$) \\
    $33,96\%$ do TítuloB ($=0,58824/(1,14379+0,58824)$)
\end{enumerate}

Há diferença entre a razão de cotas e as razões de títulos para ambos os participantes. Essa diferença se dá pelo fato da alocação de $50\%$ de cada contribuição no tempo $t=2$ comprarem proporções de títulos diferentes das proporções de títulos no tempo $t=1$, afinal os preços flutuaram de forma a não manter uma proporcionalidade entre eles no tempo.

Mas isso não traz qualquer transferência de riqueza, pois está sendo usada marcação MTM, conforme se verifica. 
Vamos calcular para Marcos.

Diferença entre razão de cotas e razão de:\\
TítuloA: $(34,48\% - 34,22\%) \times 1,6111$ títulos no plano ao preço de $0,90 = \$0,0038$.\\
TítuloB: $(33,96\% - 34,22\%) \times 1,7320 \times 0,85 = - \$0,0038$.\\
A soma é igual a $0$. \\

Para Ana, tem-se a diferença entre razão de cotas e razão de:\\
TítuloA: $(65,52\% - 65,78\%) \times 1,6111 \times 0,90 = - \$0,0038$.\\
TítuloB: $(66,04\% - 65,78\%) \times 1,7320 \times 0,85 = \$0,0038$.\\
A soma é igual a $0$. \\

Portanto, não há qualquer transferência de riqueza. Vamos verificar para HTM.

\ \

\noindent {\bf Situação HTM}

\begin{enumerate}[label=\alph*.]
    \item No tempo $t=1$, Ana contribui com \$1 comprando 1 cota.\\
    \$0,50 foi destinado a comprar TítuloA ($50\% \times 1/ 1 = 0,500$ títulos) e\\
    \$0,50 foi destinado a comprar TítuloB ($50\% \times 1/ 0,90 = 0,556$ títulos).
    \item No tempo $t=2$, o valor da cota deste plano passa a ser $1,01056$ ($=1 \times \frac{0,500 \times 1,01 + 0,556 \times 0,91}{0,500 \times 1,00 + 0,556 \times 0,90}$).
    \item No tempo $t=2$, Ana contribui com mais \$1, que compra $1/1.01056$ novas cotas ($= 0,98955$ cotas) que equivalem a:\\
    $50\% \times1/0,90$ TítuloA ($= 0,556$ títulos) e\\
    $50\% \times1/0,85$ TítuloB ($= 0,588$ títulos).\\
    Ana passa a ter $1,98955$ cotas ($= 1 + 0,98955$ cotas) que equivalem a:\\
    $1,0556$ TítuloA ($= 0,500 + 0,556$ títulos) e\\
    $1,1438$ TítuloB ($= 0,556 + 0,588$ títulos).
    \item No tempo $t=2$, Marcos contribui com \$1, que compra $1/1.01056$ cotas ($= 0,98955$ cotas) que equivalem a:\\
    $50\% \times 1/0,90$ TítuloA ($= 0,556$ títulos) e\\
    $50\% \times 1/0,85$ TítuloB ($= 0,588$ títulos).
    \item Plano possui valor total de $\$3,01$ ($1,98955 + 0,98955 = 2,97911$ cotas, a $1,01056$ o valor da cota).
    \item Ana possui $66,78\%$ das cotas do plano ($=1,98955/(1,98955+0,98955)$). Seus recursos dentro do plano são responsáveis por:\\
    $65,52\%$ do TítuloA ($=1,05556/(1,05556+0,55556)$)\\
    $66,04\%$ do TítuloB ($=1,14379/(1,14379+0,58824)$)\\
    \item Marcos possui 33,22\% das cotas do plano ($=0,98955/(1,98955+0,98956)$). Seus recursos dentro do plano são responsáveis por:\\
    $34,48\%$ do TítuloA ($=0,55556/(1,05556+0,55556)$) \\
    $33,96\%$ do TítuloB ($=0,58824/(1,14379+0,58824)$)
\end{enumerate}

Há diferença entre a razão de cotas e as razões de títulos para ambos os participantes.

Marcos, ao entrar no plano, fez com que Ana tivesse um percentual de cotas no plano maior que seus percentuais de títulos. Já Marcos está com um percentual de cotas no plano menor que seus percentuais de títulos. Como se observa, ao entrar no plano, Marcos transferiu riqueza para Ana.

Sob o ponto de vista de Marcos, o valor dessa transferência é igual a:\\
TítuloA: $(33,22\% - 34,48\%) \times 1,6111 \times 0,90 = - \$0,018$.\\
TítuloB: $(33,22\% - 33,96\%) \times 1,7320 \times 0,85 = - \$0,011$.\\
A soma é igual a $-0,029$.\\
Em termos percentuais, equivale a $2,9\%$ de sua contribuição.

Ana recebeu, em transferência de riqueza, $(66,78\% - 65,52\%) \times 1,6111 \times 0,90 + (66,78\% - 66,04\%) \times 1,7320 \times 0,85 = \$0,029$ no tempo $t=2$.\\ 
Isso equivale a $2,9\%$ de sua contribuição ou $1,46\%$ do seu saldo do plano no tempo $t=2$ (considerando o valor do plano em HTM).

\subsubsection{Exemplo 3 - parte da carteira em HTM e parte em MTM}

Equivalente ao exemplo 2, à exceção de que apenas uma parte da carteira de ativos está em HTM (por ex. 30\%, aplicado a ambos os títulos).

\begin{enumerate}[label=\alph*.]
    \item No tempo $t=1$, Ana contribui com \$1 comprando 1 cota.\\
    \$0,50 foi destinado para comprar TítuloA ($50\% \times 1/ 1 = 0,500$ títulos) e\\
    \$0,50 foi destinado a comprar TítuloB ($50\% \times 1/ 0,90 = 0,556$ títulos).
    \item No tempo $t=2$, o valor da cota deste plano passa a ser $0,94873$ ($=1 \times \big[ 30\% \frac{0,500 \times 1,01 + 0,556 \times 0,91}{0,500 \times 1,00 + 0,556 \times 0,90} + 70\% \frac{0,500 \times 0,90 + 0,556 \times 0,85}{0,500 \times 1,00 + 0,556 \times 0,90} \big]$).
    \item No tempo $t=2$, Ana contribui com mais \$1, que compra $1/0,94873$ novas cotas ($= 1,05404$ cotas) que equivalem a:\\
    $50\% \times 1/0,90$ TítuloA ($= 0,556$ títulos) e\\
    $50\% \times 1/0,85$ TítuloB ($= 0,588$ títulos).\\
    Ana passa a ter $2,05404$ cotas ($= 1 + 1,05404$ cotas) que equivalem a:\\
    $1,0556$ TítuloA ($= 0,500 + 0,556$ títulos) e\\
    $1,1438$ TítuloB ($= 0,556 + 0,588$ títulos).
    \item No tempo $t=2$, Marcos contribui com \$1, que compra $1/0,94873$ cotas ($= 1,05404$ cotas) que equivalem a:\\
    $50\% \times 1/0,90$ TítuloA ($= 0,556$ títulos) e\\
    $50\% \times 1/0,85$ TítuloB ($= 0,588$ títulos).
    \item Plano possui valor total de $\$2,95$ ($2,05404 + 1,05404 = 3,10809$ cotas, a $0,94873$ o valor da cota).
    \item Ana possui 66,09\% das cotas do plano ($=2,05404/(2,05404+1,05404)$). Seus recursos dentro do plano são responsáveis por:\\
    65,52\% do TítuloA ($=1,05556/(1,05556+0,55556)$)\\
    66,04\% do TítuloB ($=1,14379/(1,14379+0,58824)$)\\
    \item Marcos possui 33,91\% das cotas do plano ($=1,05404/(2,05404+1,05404)$). Seus recursos dentro do plano são responsáveis por:\\
    34,48\% do TítuloA ($=0,55556/(1,05556+0,55556)$) \\
    33,96\% do TítuloB ($=0,58824/(1,14379+0,58824)$)
\end{enumerate}

Há diferença entre a razão de cotas e as razões de títulos para ambos os participantes. Como se observa, ao entrar no plano, Marcos transferiu riqueza para Ana.

Sob o ponto de vista de Marcos, o valor dessa transferência é igual a:\\
TítuloA: $(33,91\% - 34,48\%) \times 1,6111 \times 0,90 = - \$0,00820$.\\
TítuloB: $(33,91\% - 33,96\%) \times 1,7320 \times 0,85 = - \$0,00069$.\\
A soma é igual a $-0.00892$.\\
Em termos percentuais, equivale a $0,89\%$ de sua contribuição no mês. Esse resultado é $30\%$ do resultado de transferência de riqueza no exemplo em que HTM foi aplicado para toda a carteira (exemplo 2 - HTM).

Ana recebeu, em transferência de riqueza, $(66,09\% - 65,52\%) \times 1,6111 \times 0,90 + (66,09\% - 66,04\%) \times 1,7320 \times 0,85 = \$0,00892$ no tempo $t=2$.\\ 
Isso equivale a $0,89\%$ de sua contribuição mensal. Esse resultado é $30\%$ do resultado de transferência de riqueza no exemplo em que HTM foi aplicado para toda a carteira (exemplo 2 - HTM). \\

Portanto, aplicar HTM em apenas uma parcela dos ativos não evita a transferência de riqueza. Além disso, a transferência de riqueza não desaparece ainda que o HTM seja aplicado apenas para um único (ou grupo restrito de) título, porque todos os títulos interferem no valor final da cota do plano.

\subsection{HTM: transferência de riqueza na saída de recursos}

Em ambas as situações, o plano se inicia com cotas tendo o mesmo valor tanto na marcação MTM quanto na HTM, a fim de que seja isolado o efeito das diferentes metodologias quando da saída de recursos do plano. Além disso, há apenas um único tipo de título e não há entradas. Como se verá, não serão necessários valores nestes exemplos. Entretanto, faremos o relacionamento com o exemplo 1 da subseção 2.3.1 também.

\ \

\noindent {\bf Situação MTM}

\begin{enumerate}[label=\alph*.]
    \item Plano tem $W$ títulos e $C$ cotas.
    \item Marcos investe $Z$ em dinheiro, comprando $X$ cotas, que equivalem a $Y$ títulos.
    \item Plano passa a ter $C + X$ cotas e $W + Y$ títulos.
    \item Após $t$ anos, Marcos retira, de forma não prevista, suas $X$ cotas, que equivalem aos $Y$ títulos, recebendo $H$ em dinheiro.
    \item Plano volta a ter $C + X - X = C$ cotas e $W + Y - Y = W$ títulos.
\end{enumerate}

Não houve transferência de riqueza na saída de recursos. Marcos resgatou o valor que havia investido com o rendimento obtido pelo título que estava na carteira do plano. O plano retornou à situação que estava antes da entrada de Marcos no que se refere a títulos e cotas.

\ \

\noindent {\bf Situação HTM}

\begin{enumerate}[label=\alph*.]
    \item Plano possui $W$ títulos e $C$ cotas.
    \item Marcos investe $Z$ em dinheiro, comprando $X$ cotas, que equivalem a $Y$ títulos.
    \item Plano passa a ter $C + X$ cotas e $W + Y$ títulos.
    \item Após $t$ anos, Marcos retira, de forma não prevista, suas $X$ cotas, que paga $G$ em dinheiro.\\
    $G$ é diferente de $H$, porque os valores de cada cota nas situações 1 e 2 são diferentes depois de $t$ anos. Nesta situação, as cotas estão avaliadas considerando o valor na curva (HTM).\\
    $G$ equivale a $Y'$ títulos. $Y'$ pode ser maior ou menor que $Y$, depende dos preços de mercado do título $Y$.
    \item Plano volta a ter $C + X - X = C$ cotas, mas passa a ter $W + Y - Y'$ títulos
\end{enumerate}

Caso $Y - Y' > 0$, Marcos teve uma perda na retirada de seus recursos. Essa perda ficou com os participantes remanescentes no plano, afinal $W + Y - Y' > W$. Houve uma transferência de riqueza de Marcos para os participantes.

Caso $Y - Y' < 0$, Marcos obteve um ganho na retirada de seus recursos às custas dos participantes remanescentes no plano, afinal $W + Y - Y' < W$. Houve uma transferência de riqueza dos participantes para Marcos.

Em ambos os casos, a transferência de riqueza é igual a $G - H$, ou seja, a diferença nos valores de saída utilizando HTM e MTM.

Ao realizar a comparação de metodologia para mensuração da transferência de riqueza de quem entra e permanece no plano (seção 2.3) com a metodologia para mensuração da transferência de riqueza de quem sai do plano (esta seção), observa-se que são congruentes como se vê abaixo.

\ \

\noindent {\bf Revisitando o exemplo 1 - HTM (subseção 2.3.1)}

\ \

Vamos retomar o Exemplo 1 (na subseção 2.3.1) no caso HTM. Considere que Ana, logo após sua contribuição no tempo $t=2$, retire seu recurso (por qualquer motivo possível).

Ana possui $1,99$ cotas que valem cada uma $1.01$. Portanto, seu resgate é $\$2,01$ ($= 1,99 \times 1,01$). Para liquidar tal saída, o plano precisa vender $2,01/0,90 = 2,33$ títulos no mercado. Como o plano possui $3,22$ títulos, sobrarão $0,989$ títulos no plano após a saída de Ana.

Entretanto, a contribuição de Marcos no tempo $t=2$ possibilitou que o plano comprasse $1,111$ títulos. Dessa forma, Marcos perdeu $0,122$ títulos ($1,111-0,989$) com a saída de Ana. Em termos monetários é igual a $\$ 0,11$ ($=0,122 \times 0,90$).

Caso a cotização fosse por meio de MTM, Ana sairia com resgate de $\$1,90$ ($2,111 \times 0,90$, referente aos títulos que seus recursos compraram mensurados pelo preço de mercado na venda). 

Ora, $2,01 - 1,90 = 0,11$ (retirada caso mensuração seja HTM $-$ retirada caso mensuração seja MTM).

Ou seja, a diferença calculada entre os valores de saídas na metodologia HTM ($\$2,01$) e na metodologia MTM ($\$1,90$) reflete a transferência de riqueza no momento da saída de Ana.

\ \ 

Para finalizar o exemplo, vamos supor agora que os preços de mercado estão mais próximos dos da curva no momento da retirada, ou seja, houve uma tentativa de se prever a retirada/realocação (aleatórios) com um título de vencimento mais próximo da data de quitação da saída.

Vamos retomar o Exemplo 1 (na subseção 2.3.1) no caso HTM. Considere que Ana, logo após sua contribuição no tempo $t=2$, retire seu recurso (por qualquer motivo possível). Ao invés do preço de mercado do título ser $0,90$, considere $0,98$, preço bem mais próximo do valor da curva ($1,01$).

No tempo $t = 2$, Ana contribui com mais \$1, que compra $1/1,01$ cotas ($= 0,99$ cotas) que equivalem a $1/0,98$ títulos ($= 1,02$ títulos), afinal o título foi adquirido no mercado no tempo $t = 2$ ao preço de $\$0,98$. Ana passa a ter $1,99$ cotas ($= 1 + 0,99$ cotas) que equivalem a $2,02$ títulos ($= 1 + 1,02$ títulos).

Portanto, sua retirada, com HTM, é $\$2,01$ ($= 1,99 \times 1,01$). O título agora vale no mercado R$\$ 0,98$ (preço mais próximo da mensuração na curva do que R$\$ 0,90$). Para liquidar tal saída, o plano precisa vender $2,01/0,98 = 2,05$ títulos no mercado. Como o plano possui $3,04$ títulos ($2,02$ de Ana e $1,02$ de Marcos), sobrarão $0,99$ títulos no plano após a saída de Ana.

Entretanto, a contribuição de Marcos no tempo $t=2$ possibilitou que o plano comprasse $1,02$ títulos. Dessa forma, Marcos perdeu $0,03$ títulos ($1,02-0,99$) com a saída de Ana. Em termos monetários é igual a $\$ 0,03$ ($=0,03 \times 0,98$).

Caso a cotização fosse por meio de MTM, Ana sairia com resgate de $\$1,98$ ($2,02 \times 0,98$, referente aos títulos que seus recursos compraram, mensurados pelo preço de mercado na venda). 

Ora, $2,01 - 1,98 = 0,03$ (retirada caso mensuração seja HTM $-$ retirada caso mensuração seja MTM).

Ou seja, a diferença calculada entre os valores de saídas na metodologia HTM ($\$2,01$) e na metodologia MTM ($\$1,98$) reflete a transferência de riqueza no momento da saída de Ana.

Esse trecho ilustra que sempre haverá transferência na saída de recurso caso o valor na curva não seja {\bf exatamente} igual ao valor de mercado. Ainda que sejam vendidos os títulos mais curtos (e já no vencimento) para quitar uma saída, ou seja, aqueles que, em tese, possuem o valor na curva mais próximo do de mercado, haverá transferência de riqueza se houver qualquer diferença de valores, ou ainda uma necessidade maior de venda de títulos para quitar uma saída contingente, além dos que venceram, ou ainda uma sobra de títulos que venceram mas não foram usados para quitar uma retirada. Não há razão para que o prejuízo destas diferenças, advindo do modelo de mensuração de ativos, seja pago pelo patrimônio dos participantes.

\section{Metodologia}

Foi programado no pacote R\footnote{Código aberto que pode ser compartilhado após solicitação para o {\it e-mail} do autor inicial.} (gratuito para {\it download}) a dinâmica de um plano CD entre Dez/2005 e Dez/2024. As variáveis consideradas foram: 

\begin{enumerate}[label=\alph*.]
    \item número de participantes inicial; 
    \item salário mensal de participação; 
    \item taxa anual de saídas (\%); 
    \item taxa anual de entradas (\%);
    \item alocação em títulos de vencimento em Dez/2025 (20 anos após a data inicial);
    \item alocação em títulos de vencimento em Dez/2030 (25 anos após a data inicial); 
    \item alocação em títulos de vencimento em Dez/2035 (30 anos após a data inicial) e
    \item quando do pagamento de uma saída do plano, os títulos que serão preferencialmente vendidos no mercado serão (i) os mais antigos adquiridos na carteira ou (ii) serão os mais recentes ou (iii) serão os de mais curto prazo.
\end{enumerate}

Os cenários elaborados para apresentação neste artigo buscam cobrir situações de planos com poucos e muitos participantes, com salários menores e maiores de participação, planos recentes (que teriam poucas saídas) ou planos maturados (com saídas mais relevantes) e planos sem crescimento e planos com crescimento mais significativo de novos participantes.
Como já afirmado, a saída do plano se refere ao pagamento de um resgate, uma portabilidade ou um benefício, qualquer que seja este. Por premissa deste estudo, essa saída é total, ou seja, todo o saldo daquele participante em específico é retirado do plano.

As entradas, saídas, cálculo de cotas, alocação dos títulos, enfim, o estudo foi todo conduzido de forma anual, pelo horizonte de 20 anos.

Para tornar as oscilações de preços de títulos e valores de cotas verdadeiras, fáticas e reais, utilizamos dados das estruturas a termo de taxa de juros de cupom de IPCA (ETTJ), de Dez/2005 a Dez/2024 (20 anos). Esses dados foram extraídos do sítio eletrônico da B3\footnote{$https://www.b3.com.br/pt_br/market-data-e-indices/servicos-de-dados/market-data/consultas/mercado-de-derivativos/precos-referenciais/taxas-referenciais-bm-fbovespa/$}. 

Foram coletadas as ETTJ (DI x IPCA) no último dia com a informação disponível de cada ano. Os prazos nesses dados estão em dias corridos. Para os casos em que foi necessário interpolar a ETTJ para algum vértice anual, foram utilizados {\it splines} cúbicos (não há qualquer prejuízo aos resultados caso seja utilizado outro método para interpolação).

\begin{figure}[ht!]
\caption{Séries temporais anuais das taxas de juros reais em IPCA (cupom de IPCA) para os prazos 5, 10 e 20 anos, de Dez/2005 a Dez/2024.}
\centering
\includegraphics[width=12cm]{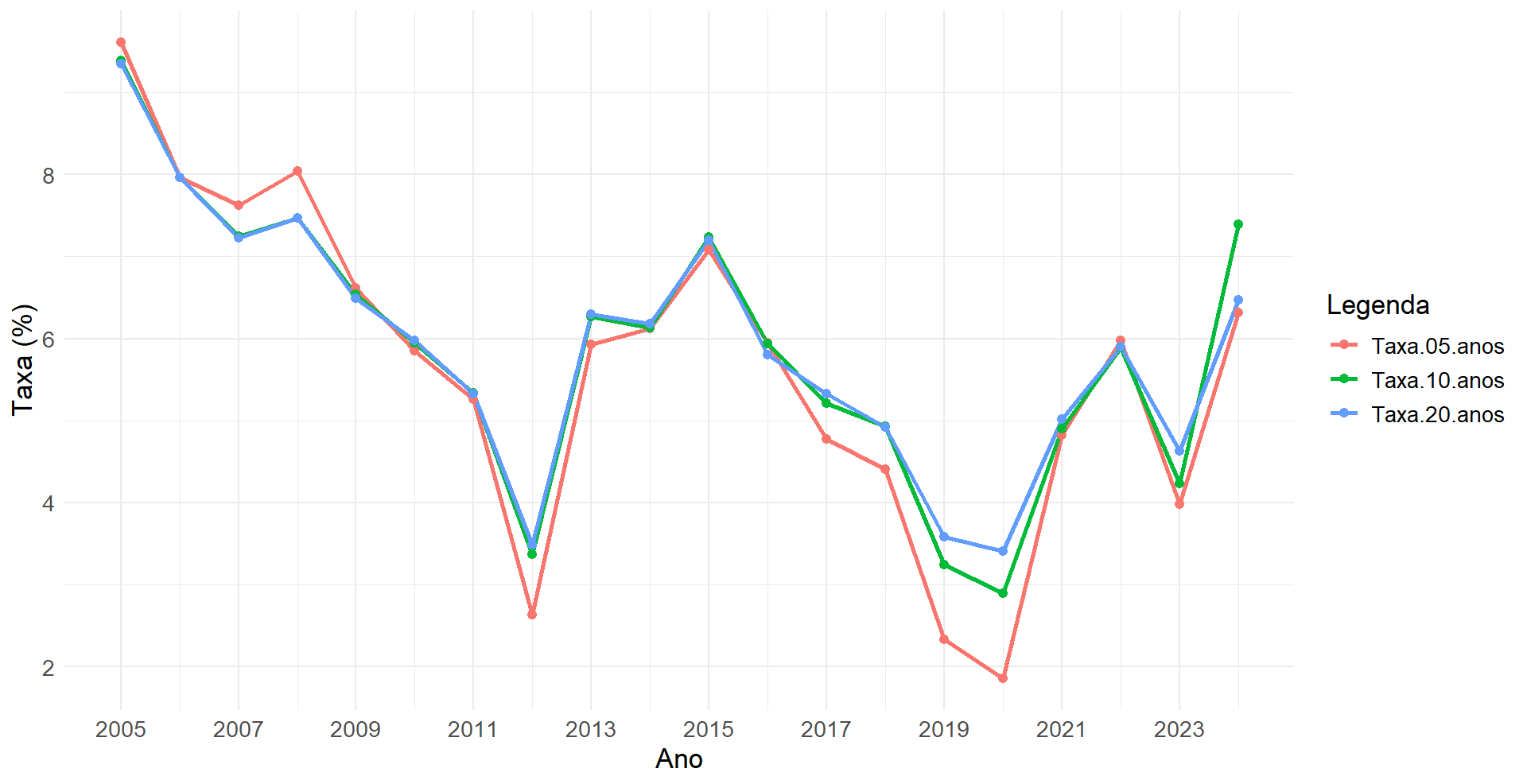}
\label{fig:ettj}
\end{figure}

A Figura \ref{fig:ettj} retrata as séries temporais das taxas de juros reais em IPCA (cupom de IPCA) ao longo dos anos, para os prazos 5, 10 e 20 anos. Observa-se que os anos de vale das taxas são 2012 e 2020. Há períodos de queda paulatina (entre 2005 e 2012 e entre 2015 e 2020). Esses marcos são importantes porque interferirão nos volumes de transferência de riqueza positiva ou negativa entre os participantes quando do uso da marcação na curva (HTM) e serão mais claros de se identificar ao se mensurar as transferências ocorridas nas saídas de recursos nestes períodos.

Sobre a alocação de ativos, optou-se por usar 3 títulos zero-cupom (com vencimentos em 20, 25 e 30 anos após a data inicial - Dez/2005, ou seja, com vencimentos em Dez/2025; Dez/2030 e Dez/2035). Os preços de mercado foram calculados com a estrutura a termo de taxa de juros de cupom de IPCA, para tornar mais próximo da realidade de planos CD brasileiros, que costumam investir em NTN-B (Notas do Tesouro Nacional com fluxos de caixa corrigidos pelo IPCA). O fato de o estudo ter sido conduzido com títulos zero-cupom ao invés de títulos com pagamento de cupons não prejudica as conclusões a respeito da identificação e mensuração de transferência de riqueza. Além disso, uma das justificativas que se observa em matérias jornalísticas\footnote{Revista da Previdência Complementar, ano 44, n. 456 - Jan/Fev 2025 - https://www.agenciawiser.com.br/revista-previdencia/} a respeito da adoção recente da mensuração na curva (HTM) nos planos CD é o prazo longo dos ativos de renda fixa. Como se observará nos resultados, quanto mais longos os títulos, maiores serão as transferências de riqueza, caso adotada marcação na curva (HTM). 

Uma vez definida a alocação (\%) dos ativos em um determinado cenário do estudo, essa alocação será aplicada quando da entrada de novas contribuições a cada ano e também será aplicada quando do cálculo das saídas. Por exemplo, caso a alocação seja [0\%; 33\%; 67\%] nos títulos com vencimento em Dez/2025, Dez/2030 e Dez/2035, respectivamente, as saídas serão pagas com a alienação destas proporções de títulos, assim como as novas contribuições serão investidas nos títulos com base nessas proporções.

Por fim, na mensuração dos ativos pelo HTM, faz diferença o pagamento das saídas ser feito com base na venda dos títulos mais antigos na carteira do plano, ou os mais recentes ou os de prazos mais curtos. Dessa forma, o estudo também permite estabelecer um desses 3 critérios para venda dos títulos para pagamento das saídas.

A transferência de riqueza foi calculada: \\
(i) para cada um dos participantes que permanecem no plano até Dez/2024; e\\
(ii) para cada participante que eventualmente saia do plano nesses 20 anos.

Com relação ao item (i), o valor mensurado de transferência de riqueza para cada participante que permanece no plano acaba sendo cumulativo para o período. Ou seja, caso tenha havido, em um determinado ano, uma transferência de riqueza que prejudicou algum participante e, em um ano posterior, tenha havido uma transferência de riqueza que o tenha favorecido, esses valores estarão refletidos, no último ano, na razão de cotas e na razão de títulos referentes às contribuições feitas por aquele participante ao longo do tempo.

Portanto, caso haja uma compensação de transferências de riqueza no tempo para cada participante, isso está sendo levado em consideração no estudo. Como visto nos resultados, isso não anula transferências de riqueza nem para participantes que permanecem nem para os que retiram recursos.

Sobre o item (ii), a respeito do cálculo da transferência de riqueza para cada participante que eventualmente saia do Plano, utilizamos a diferença de saída caso o plano estivesse marcado na curva (HTM) e a mercado (MTM), como já explicado na seção 2.

Abaixo descrevemos o algoritmo utilizado para identificar e mensurar transferências de riqueza caso seja adotada a marcação na curva (HTM) em um plano CD.

\ \

\noindent {\bf Algoritmo:}

\begin{enumerate}
    \item No momento inicial (Dez/2005, $t = 1$), valor da cota MTM = Valor da cota HTM = 1.
    \item Cálculo do número de cotas adquiridas, em $t=1$, para cada participante na massa inicial, tanto para MTM quanto para HTM, com base em suas contribuições em Dez/2005.
    \item Cálculo do número (alocação) de títulos totais do plano por vencimento de título no tempo $t=1$. Para MTM e HTM, essa alocação será calculada a cada tempo $t$. Para HTM, além disso, a alocação é controlada também pelo ano de aquisição dos títulos ($k=1,...,t$).
    \item Cálculo do número de títulos subjacentes por cada participante, no tempo $t=1$, por vencimento do título.
    \item Para os tempos $t=2,...,20$:
        \begin{enumerate}[label=\alph*.]
            \item Cálculo do valor da cota MTM no tempo $t$, usando a cota MTM do tempo anterior $t-1$ a alocação de títulos no tempo $t-1$ e os preços de mercado dos títulos nos tempos $t-1$ e $t$. \\
            
            Cálculo do valor da cota HTM no tempo $t$, usando a cota HTM do tempo anterior $t-1$ a alocação de títulos no tempo $t-1$, com controle pelo ano de aquisição, e os preços na curva dos títulos nos tempos $t-1$ e $t$, com controle dos preços pelo ano de aquisição (curva de cada título varia conforme data de aquisição).
            
            \item Caso tenha saídas, indicação dos registros dos participantes que sairão (sorteio daqueles que estavam no plano no tempo $t-1$ com base no \% de saídas).
            
            \item Número de cotas de cada participante no tempo $t$ (para MTM e para HTM, separadamente):\\
            
            Para os participantes que não saíram: número de cotas de cada participante no tempo $t$ é igual a do tempo $t-1$ somado às cotas adquiridas com contribuição no tempo $t$ e valor da cota no tempo $t$.\\
            
            Para os participantes que saíram: número de cotas desses participantes é zerada no tempo $t$.
            
            \item Número de títulos subjacentes de cada participante no tempo $t$ para os vencimentos distintos:\\
            
            Para os participantes que não saíram: número de títulos de cada participante no tempo $t$ para os vencimentos distintos é igual a do tempo $t-1$ somado aos títulos adquiridos com contribuição no tempo $t$.\\
            
            Para os participantes que saíram: número de títulos de cada participante no tempo $t$ é zerado.
            
            \item Cálculo do valor de saída dos participantes que saíram (para MTM e para HTM separadamente):\\
            
            Valor de saída de cada participante que sai = número de cotas do participante no tempo $t-1$ multiplicado pelo valor da cota no tempo $t$ (MTM ou HTM). A contribuição do ano $t$ não entra para ser imediatamente retirada.
            
            \item Adição dos participantes que entram no plano no tempo $t$, com cálculo do número de cotas de cada um e número de títulos de cada um para os diferentes vencimentos.
            
            \item Alocação de títulos do plano no tempo $t$:\\
            
            Para MTM: alocação de títulos no tempo $t-1$ adicionada à alocação das contribuições feitas no tempo $t$ (tanto dos participantes remanescentes quanto dos novos entrantes) nos títulos com base no preço de mercado no tempo $t$, subtraída da alocação (número de títulos) necessária para pagar as saídas no tempo $t$.\\
            
            Para HTM: \\ 
            
            (i) alocação de títulos no tempo $t-1$, separada por ano de aquisição (até o tempo $t-1$), é mantida para o tempo $t$. \\
            A alocação dos títulos adquiridos no ano $t$ no tempo $t$ é calculada com base nas contribuições (tanto do participante antigos quanto dos novos entrantes) convertidas nos títulos com base no preço de mercado no tempo $t$. \\
            
            (ii) cálculo dos títulos necessários para a saída com base no preço de mercado dos títulos no tempo $t$.\\

            (iii) para abater estes títulos, calculados no item anterior, da alocação do plano (em HTM), depende se a estratégia é vender os títulos mais antigos no plano ou os mais recentes ou os mais curtos.\\
            Caso a estratégia seja sobre títulos mais antigos, os títulos serão subtraídos da alocação do mais antigo (adquiridos em $t=1$) até o mais recente, até quitar todas as saídas.\\
            Caso a estratégia seja sobre títulos mais recentes, os títulos serão subtraídos da alocação do mais recente (adquiridos em $t=t-1$) até os mais antigos, até quitar todas as saídas.\\
            Caso a estratégia seja sobre títulos mais curtos, os títulos serão subtraídos da alocação do mais curto (adquiridos em $t=t-1,...,1$, do mais recente ao mais antigo) até os mais longos (adquiridos em $t=t-1,...,1$, do mais recente ao mais antigo), até quitar todas as saídas.

            \item Com as alocações em $t$ tanto para mensuração MTM quanto HTM, é possível calcular o valor da cota em $t+1$ (item 5a) e o processo é refeito até $t=20$.
        \end{enumerate}

        \item Cálculo da perda ou do ganho na saída decorrente da transferência de riqueza para cada participante que sai a cada tempo $t=2,...,20$:\\
        
        Cálculo da diferença entre o valor de saída considerando MTM e o valor de saída considerando HTM. Cálculo desta diferença em termos \% em relação ao valor de saída considerando MTM.\\

        \item Cálculo da perda ou do ganho para cada participante que permaneceu no plano até Dez/2024 ($t=20$) decorrente da transferência de riqueza:\\
        
        Cálculo da razão de cotas (HTM) para cada participante no tempo $t=20$.\\
        Cálculo da razão de títulos para cada participante no tempo $t=20$.\\
        Aplica a diferença dessas razões na alocação de títulos do plano no tempo $t=20$ multiplicado pelo preço de mercado destes títulos no tempo $t=20$. Cálculo deste montante em termos \% em relação ao valor total do plano no MTM no tempo $t=20$.

\end{enumerate}

\section{Resultados}

Essa seção de resultados se divide em 3 subseções: 

\begin{enumerate}
    \item Casos exemplificativos - evolução de plano no período de análise: HTM x MTM:\\
    (a) sem entradas e sem saídas e investimento em um único título com vencimento coincidindo com o fim do período do estudo;\\
    (b) plano com saída que gere uma transferência de riqueza tornando o plano insustentável financeiramente para os remanescentes.
    \item Resultados de transferências de riqueza na saída de recursos.
    \item Resultados de transferências de riqueza na entrada de recursos.
\end{enumerate}

\ \ 

Os resultados apresentados nesta seção foram obtidos considerando que a carteira de ativos toda foi marcada na curva. Caso uma parcela da carteira esteja marcada na curva, uma redução proporcional no valor de transferência de riqueza será obtido. A transferência só será eliminada caso não haja parcela em HTM.

Para os itens 2 e 3 acima, foram construídas as seguintes combinações de cenários:

\begin{enumerate}[label=\alph*.]
    \item Número de participantes no plano: $1.000$, $10.000$ ou $50.000$.
    \item Salário mensal de participação: $\$5.000$ ou $\$15.000$ (alíquota total de $15\%$ e $13$ pagamentos por ano).
    \item Taxa anual de saídas: $0,10\%$, $3,57\%$, $7,03\%$ ou $10,5\%$ (valores definidos de forma uniforme entre $0,10\%$ e $10,5\%$).
    \item Taxa anual de entradas: $0\%$, $3,33\%$, $6,67\%$ ou $10\%$.
    \item Venda dos títulos na carteira para pagamento em caso de saída: antigos, recentes, curtos.
    \item Alocação em títulos de vencimento em Dez/2025: $0\%$, $25\%$, $50\%$, $75\%$ ou $100\%$.
    \item Alocação em títulos de vencimento em Dez/2030: $0\%$, $25\%$, $50\%$, $75\%$ ou $100\%$.
    \item Alocação em títulos de vencimento em Dez/2035: $0\%$, $25\%$, $50\%$, $75\%$ ou $100\%$. 
\end{enumerate}

Para as alocações em títulos, não foi permitida alocação negativa, ou seja, todas as alocações foram maiores ou iguais a zero e a soma das alocações sempre foi igual a $100\%$. Portanto, definidas as alocações em títulos com vencimento em Dez/2025 (curto) e Dez/2030 (médio) automaticamente está definida a alocação no título com vencimento em Dez/2035 (longo). Esses cenários geram $4.320$ combinações.

Sobre os cenários de taxa de entrada, estes buscam ilustrar situações de planos fechados a novos participantes (taxa igual a zero) e planos com muitas entradas ($10\%$). Sobre as taxas de saída, os cenários buscam cobrir planos com poucas saídas (taxa igual a $0,10\%$, afinal taxa igual a zero é inverossímil porque as saídas podem ocorrer por morte, invalidez, aposentadoria, perda do vínculo etc., não estando sob poder de decisão do plano, além de realocações que impliquem vendas de títulos) e planos com saídas mais relevantes ($10,5\%$). O escalonamento entre as taxas mínima e máxima de saída, mínima e máxima de entrada, mínima e máxima de alocação nos ativos foi uniforme.

\subsection{Casos exemplificativos - evolução de plano no período de análise: HTM x MTM}

Um plano que investiu em um único título quando da entrada de participantes no momento inicial e que usa mensuração de ativos na curva (HTM) terá, na data do vencimento deste título, o mesmo valor acumulado de um plano que usa mensuração a mercado (MTM), sem novas entradas e sem qualquer saída no período em ambos, ou seja, sem possibilidades de transferência de riqueza no plano com uso da marcação HTM.

\begin{figure}[ht!]
\caption{Séries temporais dos planos utilizando MTM e HTM para mensuração dos ativos, considerando 2 participantes, início em Dez/2005, com contribuição anual de \$500 cada um, sem entradas e sem saídas, e com investimento todo feito em títulos com vencimento em Dez/2024.}
\centering
\includegraphics[width=14cm]{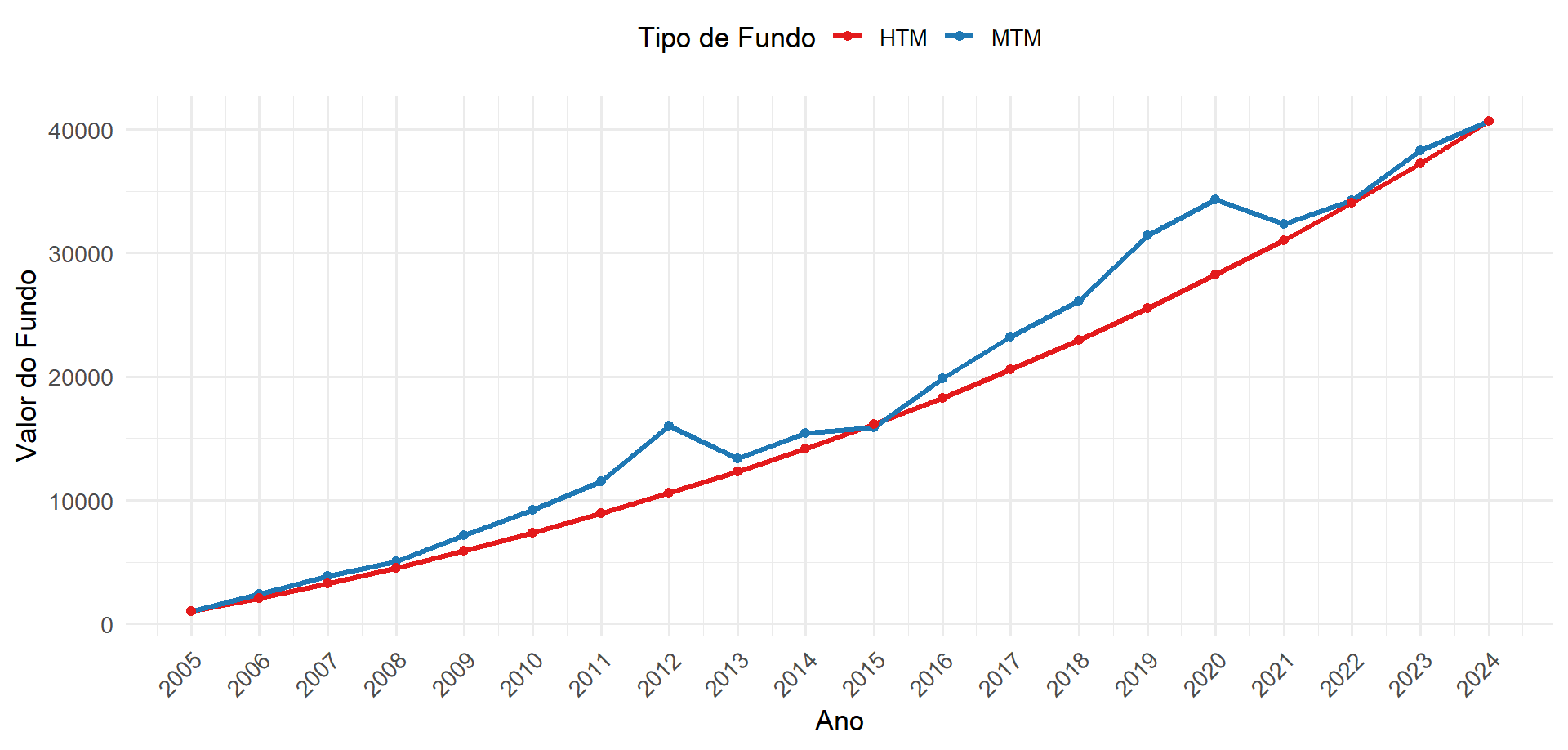}
\label{fig:evol}
\end{figure}

Para testar essa hipótese, simulamos a situação e fizemos o gráfico da evolução destes planos ao longo dos anos na Figura \ref{fig:evol}. Foram considerados 2 participantes, início em Dez/2005, com contribuição anual de \$500 cada um, sem entradas e sem saídas, e com investimento feito exclusivamente em títulos com vencimento, excepcionalmente para este exemplo, em Dez/2024 (término do nosso período de estudo).

Observamos, ao analisar de forma conjunta os gráficos das Figuras \ref{fig:ettj} e \ref{fig:evol}, que em momentos de baixa da taxa de juros (por ex. 2012), o valor dos ativos do plano por MTM sobe e fica acima do que por HTM e, quando há um aumento das taxas de juros, o movimento inverso ocorre. Considerando essa situação, uma retirada de recursos do plano (que use HTM como marcação dos ativos) por um participante em 2012, por exemplo, geraria uma transferência de riqueza sua para os que permanecerem. Afinal, seu valor de saída do plano seria inferior ao valor caso este estivesse adotando MTM. Neste exemplo hipotético foi usado um título que vence em Dez/2024, que é o último ano de análise. Dessa forma, naquela data, os planos se equivalem em valor total, pois não houve saídas ao longo do período. Como os 2 participantes contribuem com o mesmo valor em cada um dos anos, inclusive entraram de forma conjunta, também não há transferência de riqueza na entrada dos recursos: ambos mantêm sua razão de cotas igual à razão de títulos subjacentes em cada momento.

Na próxima figura, procuramos construir um cenário em que a marcação na curva (HTM) leva à insolvência do plano. Neste cenário, consideramos 2 participantes no início em Dez/2005, com contribuição única de $\$50$ cada um. Em 2012, um novo participante entra no plano, com uma única contribuição de $\$1.000$. Esse participante retira seus recursos em 2015 (ficou 3 anos no plano). Não há quaisquer entradas e nem saídas adicionais. Investimento todo feito em títulos com vencimento em Dez/2035 (o mais longo de nosso estudo).

\begin{figure}[ht!]
\caption{Séries temporais dos saldos dos planos utilizando MTM (linha azul) e HTM (linha preta) para mensuração dos ativos. Série temporal do número de cotas do Participante \#1 (que é igual ao do Participante \#2) - linha tracejada vermelha. Neste cenário, consideramos 2 participantes no início em Dez/2005, com contribuição única de $\$50$ cada um. Em 2012, um novo participante entra no Plano, com uma única contribuição de $\$1.000$. Esse participante retira seus recursos em 2015. Não há quaisquer entradas e nem saídas adicionais. Investimento todo feito em títulos com vencimento em Dez/2035.}
\centering
\includegraphics[width=14cm]{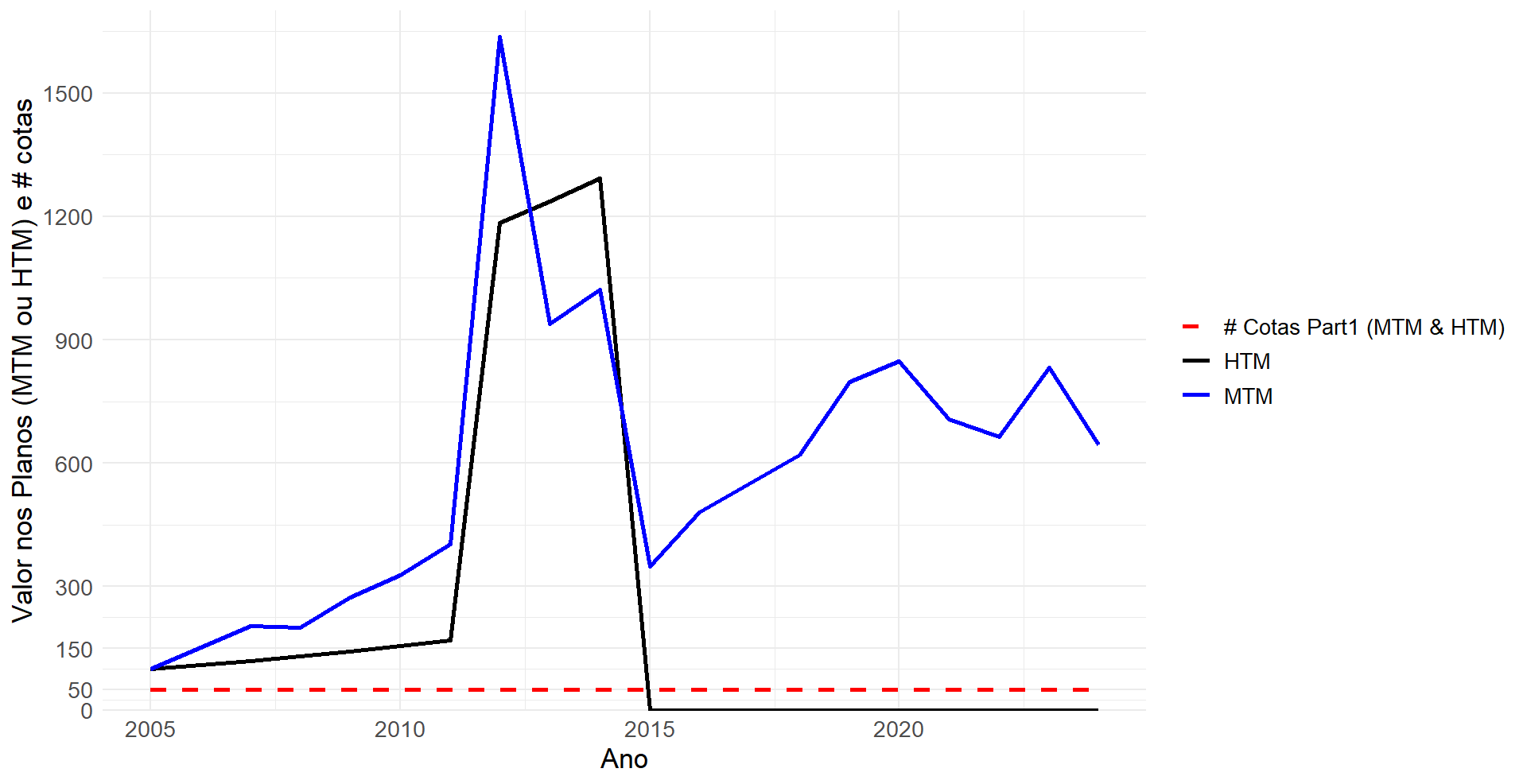}
\label{fig:evol_insolv}
\end{figure}

O Figura \ref{fig:evol_insolv} retrata a insolvência de um Plano que utiliza marcação na curva (HTM). Dois participantes entram no tempo $t=1$, com contribuição única (de $\$50$) que compra $50$ cotas cada um (linha vermelha tracejada no gráfico), tanto no cenário de plano com marcação MTM quanto no de marcação HTM (o valor da cota inicial, em Dez/2005, é igual a $1$ em ambos os casos). Como não há novas contribuições desses participantes, eles mantêm 50 cotas, cada um, até Dez/2024. Esses $100$ em 2005 compram $1,401$ títulos que vencem em 2035.

Em 2012, o participante \#3 entra no plano. Esse participante eventualmente possui uma maior capacidade financeira que os demais que estavam no plano, por isso possui uma contribuição maior. No cenário de plano com marcação HTM, comprando $540,08$ cotas com $\$ 1.000$ de contribuição. No cenário de plano com marcação MTM, comprando $156,90$ cotas com os mesmos $\$ 1.000$ de contribuição. Esses $1.000$ em 2012 compram $2,199$ títulos que vencem em 2035. Neste ano, o plano passa a ter $1,401 + 2,199 = 3,600$ títulos no total.

Em 2015, o participante \#3 retira seus recursos. Em plano com MTM, são retirados $156,90 \times 3,488862=\$ 547,41$, que equivalem exatamente a $2,199$ títulos a $\$248,95$ preço de mercado do título. Em plano HTM, são retirados $540,08 \times 2.108057 = \$1.138,54$. Em 2015, esse valor de retirada demandaria $4,5734$ títulos. Mas o plano possui apenas $3,600$ títulos (que perfazem $\$ 896,30$ em 2015). Nem a retirada completa, pelo HTM, é possível pagar ao participante \#3.

Como se pode ver pela Figura \ref{fig:evol_insolv}, o plano com marcação HTM está sem recursos financeiros a partir de 2015. Embora haja dois participantes com $50$ cotas cada um, não há mais quaisquer títulos (ou ativos) no plano. O exemplo foi montado de forma a retratar na prática outra consequência grave possível da transferência de riqueza. Observem que o participante \#3, eventualmente com maior instrução financeira para entender transferências de riqueza, aporta contribuição em 2012 (momento em que as taxas de juros caíram muito) e retira seus recursos em 2015 (momento em que as taxas de juros subiram muito). Essa estratégia montada de forma proposital por este participante procura aumentar seus ganhos com transferências de riqueza de duas formas: na entrada e na saída. O gráfico também ilustra, em um exemplo hipotético, mas não irreal, que transferências de riqueza podem também, no limite e hipoteticamente, levar à insolvência de um plano CD. 

Essa situação demanda certa flexibilidade ao participante \# 3. E há mais flexibilidade para um participante auto patrocinado, ou seja, aquele que permanece no plano, sem vínculo com o patrocinador, mas com maior liberdade sobre contribuições e retiradas ou, ainda, participantes de planos instituídos (que também possuem mais flexibilidade sobre contribuições e retiradas). Em casos especiais, como esses, participantes com maiores capacidade e instrução financeiras podem obter ganhos por transferência de riqueza às custas de participantes com menores capacidade ou instrução financeiras.

Uma liberdade para aporte e retirada de recursos facultativos, incluindo portabilidades de outros planos, aumenta em muito a possibilidade de arbitragem para ganhos com transferência de riqueza às custas dos demais participantes. Tal fato reforça que participantes com maiores capacidade e instrução financeiras se beneficiariam em prejuízo daqueles com menores capacidade ou instrução financeiras.

\subsection{Resultados de transferências de riqueza na entrada de recursos}

Nesta seção apresentamos os resultados de transferências de riqueza na entrada de recursos. A Tabela \ref{tab:tranentra} mostra os resultados para alguns cenários de planos com número inicial de participantes de $10.000$, salário de contribuição de $\$15.000$ e, para pagamento de saídas, alienam-se os títulos mais antigos na carteira (primeiro painel) e os mais curtos (segundo painel). Os valores de ganho e perda percentuais foram calculados sobre o valor do saldo acumulado em Dez/2024 na marcação a mercado (MTM).

\begin{table}[!ht]
\centering
\begin{tabular}{ c|c|c|c|c|c }
\hline   
\multicolumn{3}{c|}{Cenário} & Perda & Ganho  & Perda \\ \cline{1-3}
Alocação & Taxa & Taxa & Máx. & Máx. & (Ganho) \\ 
\multicolumn{1}{c|}{Tít.$[2025; 2030; 2035]$} & \multicolumn{1}{c|}{saída} & \multicolumn{1}{c|}{entrada} & \multicolumn{1}{c|}{(\%)} & \multicolumn{1}{c|}{(\%)} & \multicolumn{1}{c}{Médio (\%)} \\ \hline

\multicolumn{6}{c}{Estratégia de venda de títulos mais antigos} \\ \hline

$[100\%; 0\%; 0\%]$ & $0,10\%$ & $3,33\%$ & 1,93 & 10,11 & (1,84) \\ 
$[0\%; 100\%; 0\%]$ & $0,10\%$ & $3,33\%$ & 4,01 & 17,19 & (3,31) \\ 
$[0\%; 0\%; 100\%]$ & $0,10\%$ & $3,33\%$ & 8,29 & 23,97 & (4,50) \\  \hline

$[0\%; 100\%; 0\%]$ & $3,56\%$ & $3,33\%$ & 2,89 & 18,65 & (3,72) \\ 
$[0\%; 100\%; 0\%]$ & $7,03\%$ & $3,33\%$ & 3,17 & 19,31 & (4,04) \\ 
$[0\%; 100\%; 0\%]$ & $10,5\%$ & $3,33\%$ & 3,40 & 19,58 & (4,34) \\ \hline

$[0\%; 100\%; 0\%]$ & $0,10\%$ & $0,00\%$ & 0    & 0     & 0      \\ 
$[0\%; 100\%; 0\%]$ & $0,10\%$ & $6,67\%$ & 5,48 & 15,00 & (4,01) \\ 
$[0\%; 100\%; 0\%]$ & $0,10\%$ & $10,0\%$ & 7,88 & 13,07 & (3,53) \\  \hline

$[25\%; 25\%; 50\%]$ & $0,10\%$ & $3,33\%$ & 5,63 & 18,40 & (3,46) \\  \hline \hline

\multicolumn{6}{c}{Estratégia de venda de títulos mais curtos} \\ \hline

$[25\%; 25\%; 50\%]$ & $0,10\%$ & $3,33\%$ & 5,78  & 18,31 & (3,43) \\ 
$[25\%; 25\%; 50\%]$ & $3,56\%$ & $3,33\%$ & 8,02  & 16,58 & (2,87) \\ 
$[25\%; 25\%; 50\%]$ & $7,03\%$ & $3,33\%$ & 10,08 & 15,21 & (2,52) \\ 
$[25\%; 25\%; 50\%]$ & $10,5\%$ & $3,33\%$ & 11,97 & 14,73 & (2,24)\\ \hline \hline

\end{tabular}
\caption{\label{tab:tranentra} Transferência de riqueza nas entradas de recursos para os participantes que permanecem no plano até Dez/2024. Tabela contendo: (i) Perda máxima de 1 participante que permaneça no plano até Dez/2024, devido à transferência de riqueza ocorrida na mensuração HTM; (ii) Ganho máximo de 1 participante que permaneça no plano até Dez/2024, devido à transferência de riqueza ocorrida na mensuração HTM; (iii) Perda (Ganho) médio dos participantes que permaneceram no plano até Dez/2024, devido à transferência de riqueza ocorrida na mensuração HTM. Número inicial de participantes = 10.000. Salário de contribuição = \$15.000.}
\end{table}

Os valores de perdas, mostrados na Tabela \ref{tab:tranentra}, que um participante pode ter ao permanecer em um plano que utilize a marcação na curva (HTM) são significativos mas não são elevados como os valores para os participantes que saem, porque foi feito um corte em Dez/2024 e calculadas essas perdas/ganhos nessa data final. Obviamente, em algum momento esses participantes irão sair e poderão experimentar perdas/ganhos significativos.

A presença de títulos mais longos na carteira parece ampliar perdas e ganhos, máximos e médios, de participantes em função de transferências de riqueza ao permanecer no plano, afinal o valor de mercado do título longo pode ficar mais distante do valor na curva do que para títulos mais curtos. 

O aumento da taxa de saída parece afetar de forma marginal as perdas/ganhos máximos e médios de quem permanece. Essa variável possui relação positiva amplificando a perda/ganho nos resultados. Como as taxas de juros, de forma bem geral, estão caindo no período, isso justifica o fato de gerar ganhos de transferência de riqueza para aqueles que permanecem em um plano com marcação na curva (HTM) mais relevantes que as perdas. O aumento de saídas amplifica esse ganho, ao gerar perdas para os participantes que saem, com transferência de riqueza para quem permanece.

A taxa de entrada parece ter relação positiva com a perda e ganho máximos e negativa com o ganho médio. Sobre a relação com o ganho médio, pode ser devido à diluição da média em função da entrada de novos participantes com pouco tempo de acumulação no plano. Em relação às perdas e ganhos máximos, pode ser devido ao fato de entrarem mais pessoas e haver mais casos de perdas e ganhos, elevando o valor máximo encontrado.

Quando a taxa de entrada é igual a zero, a transferência de riqueza na entrada de recursos também é zero. Isso ocorre porque em nosso estudo, consideramos que todos os participantes iniciam juntos em Dez/2005 e com exatamente os mesmos valores de aporte em todos os momentos até Dez/2024. Caso os participantes efetuassem aportes com valores distintos ao longo do tempo, ocorreriam transferências de riqueza na entrada, ainda que a taxa de novos entrantes fosse igual a zero.

Sobre o painel com a estratégia de venda de títulos mais curtos, essa venda deixa na carteira os de maior prazo. Esses são os títulos com diferença relevante entre preço de mercado e valor na curva. A entrada de recursos no plano com este cenário aumenta as transferências de riqueza para quem permanece até Dez/2024.

Gráficos com alguns resultados desses experimentos estão no apêndice do artigo.

\subsection{Resultados de transferências de riqueza na saída de recursos}

Nesta seção, apresentaremos os resultados das transferências de riqueza ocorridas na saída de recursos. Como já afirmado, é muito pouco provável que um plano CD não tenha saídas em um horizonte de tempo. As saídas podem ser por morte, invalidez, aposentadoria (via retiradas parciais) ou perda do vínculo, com portabilidade ou resgate. Além disso, pode incluir venda de títulos em função de realocação de investimento. Escolhemos um conjunto limitado de cenários para poder explorar algumas diferenças nos resultados observados em uma tabela. A Tabela \ref{tab:transaida} mostra os resultados da (i) perda máxima de 1 participante (em termos \%) de todas as saídas de participantes no período de 20 anos em conjunto com o ano que foi observada essa perda máxima; (ii) ganho máximo de 1 participante (\%) de todas as saídas de participantes no período de 20 anos em conjunto com o ano que foi observado esse ganho máximo e (iii) a perda (ganho) médio dessas saídas (\%) em virtude de transferência de riqueza. Os valores de ganho e perda percentuais foram calculados sobre o valor da saída de cada participante caso o plano adotasse a marcação a mercado (MTM).

Ainda que tenhamos apresentado a perda (ganho) médio, entendemos que essa métrica não é a mais adequada. Afinal, uma média poderia ser eventualmente pequena, mas com um grupo de participantes com perdas enormes e outro grupo de participantes com ganhos igualmente enormes. Por esse motivo, também foram construídos histogramas (Figura \ref{fig:saida_hist}) com a distribuição das perdas (ganhos) por participante.

\begin{table}[!ht]
\centering
\begin{tabular}{ c|c|c|c|c|c }
\hline   
\multicolumn{3}{c|}{Cenário} & Perda & Ganho  & Perda \\ \cline{1-3}
Alocação & Taxa & Taxa & Máx. & Máx. & (Ganho) \\ 
\multicolumn{1}{c|}{Tít.$[2025; 2030; 2035]$} & \multicolumn{1}{c|}{saída} & \multicolumn{1}{c|}{entrada} & \multicolumn{1}{c|}{(\%)} & \multicolumn{1}{c|}{(\%)} & \multicolumn{1}{c}{Médio (\%)} \\ \hline

\multicolumn{6}{c}{Estratégia de venda de títulos mais antigos} \\ \hline

$[100\%; 0\%; 0\%]$ & $0,10\%$ & $3,33\%$ & 38,76 (2012) & 42,58 (2013) & 9,55 \\ 
$[0\%; 100\%; 0\%]$ & $0,10\%$ & $3,33\%$ & 49,34 (2012) & 63,05 (2013) & 13.49 \\ 
$[0\%; 0\%; 100\%]$ & $0,10\%$ & $3,33\%$ & 57,98 (2012) & 86,29 (2013) & 16,47 \\  \hline

$[0\%; 100\%; 0\%]$ & $3,56\%$ & $3,33\%$ & 49,41 (2012) & 62,98 (2013) & 16,52 \\ 
$[0\%; 100\%; 0\%]$ & $7,03\%$ & $3,33\%$ & 49,46 (2012) & 62,91 (2013) & 18,27 \\ 
$[0\%; 100\%; 0\%]$ & $10,5\%$ & $3,33\%$ & 49,52 (2012) & 62,85 (2013) & 19,58 \\ \hline

$[0\%; 100\%; 0\%]$ & $0,10\%$ & $0,00\%$    & 49,26 (2012) & 2,94 (2024)  & 20,45 \\ 
$[0\%; 100\%; 0\%]$ & $0,10\%$ & $6,67\%$    & 49,43 (2012) & 62,87 (2013) & 8,91 \\ 
$[0\%; 100\%; 0\%]$ & $0,10\%$ & $10,0\%$    & 49,52 (2012) & 62,69 (2013) & 6,88 \\  \hline 

$[25\%; 25\%; 50\%]$ & $0,10\%$ & $3,33\%$ & 52.16 (2012) & 70.16 (2013) & 14.26 \\  \hline \hline

\multicolumn{6}{c}{Estratégia de venda de títulos mais curtos} \\ \hline

$[25\%; 25\%; 50\%]$ & $0,10\%$ & $3,33\%$ & 52,16 (2012) & 70.17 (2013) & 14,21 \\ 
$[25\%; 25\%; 50\%]$ & $3,56\%$ & $3,33\%$ & 51,99 (2012) & 70,86 (2013) & 16,11 \\ 
$[25\%; 25\%; 50\%]$ & $7,03\%$ & $3,33\%$ & 51,89 (2012) & 71,24 (2013) & 17,48 \\ 
$[25\%; 25\%; 50\%]$ & $10,5\%$ & $3,33\%$ & 51,91 (2012) & 71,93 (2013) & 18,64 \\ \hline \hline

\end{tabular}
\caption{\label{tab:transaida} Transferência de riqueza na saída de recursos. Tabela contendo: (i) Perda máxima de 1 participante que saia do plano, devido à transferência de riqueza ocorrida na mensuração HTM (e ano de ocorrência dessa perda máxima); (ii) Ganho máximo de 1 participante que saia do plano, devido à transferência de riqueza ocorrida na mensuração HTM (e ano de ocorrência desse ganho máximo); (iii) Perda (Ganho) médio dos participantes que saíram do plano, devido à transferência de riqueza ocorrida na mensuração HTM. Número inicial de participantes = 10.000. Salário de contribuição = \$15.000.}
\end{table}

Primeira observação relevante são os significativos valores de perdas que um participante pode ter ao retirar seus recursos em um plano que utilize a marcação na curva (HTM). Dependendo do momento da ETTJ, essa perda, em nossos resultados, pode chegar a quase $60\%$ do seu saldo (valores máximos). Títulos mais longos na carteira parecem ampliar perdas e ganhos, máximos e médios, de participantes em função de transferências de riqueza, afinal o valor de mercado do título longo pode ficar mais distante do valor na curva do que para títulos mais curtos. Ressalta-se que ganhos são obtidos às custas de perdas de outros participantes e vice-versa. O aumento da taxa de saída parece não afetar as perdas/ganhos máximos, mas essa variável possui relação positiva com a perda média nos resultados. A taxa de entrada também parece não interferir em perdas/ganhos máximos, mas possui relação inversa com a perda média. Altas taxas de entrada aumentam, no estudo, a frequência absoluta de saídas de pessoas com poucos recursos (porque aumenta o número de participantes no plano que entraram já com o plano iniciado). Os montantes de transferência de riqueza nas saídas desses participantes, em específico, acabam diluindo a média da perda por transferência de riqueza na saída.

O cenário em que a taxa de entrada é igual a zero retrata a situação em que só há os participantes que iniciaram no plano em Dez/2005. Dessa forma, não há cenário de participante que entre em um momento ótimo para aporte de recursos a fim de obter ganho com a transferência de riqueza vinda dos demais. Ao contrário, quando há anualmente a entrada de novos participantes (taxa de entrada superior a 0\%), sempre haverá, no estudo, um participante entrando no momento ótimo para obter ganho de transferência de riqueza e, posteriormente, podendo sair, também em um momento ótimo do período para maximizar o ganho por transferência de riqueza na saída (entrar com a taxa de juros baixa e sair com a taxa de juros alta).

No que se refere ao uso da estratégia de alienação de títulos mais curtos para quitar saídas, ela tende a gerar menor transferência de riqueza na saída, mas ainda assim com valores relevantes nesse estudo. Nesse estudo, o título mais curto é o de vencimento em Dez/2025, então, por alguns anos do estudo, esse título ainda possui maturidade significativa. De qualquer forma, como exposto no exemplo da seção 2, qualquer diferença entre preço de mercado e valor mensurado na curva gerará transferência de riqueza.

Gráficos com alguns resultados desses experimentos estão no apêndice do artigo.

\subsection{Debate sobre os resultados}

A análise dos exemplos da seção 4.1 e resultados das seções 4.2 e 4.3 indica que a transferência de riqueza na saída pode ser mitigada com a tentativa de imunização do fluxo de caixa de saída, por meio da aquisição de títulos que tenham fluxos de caixa que vençam exatamente nas datas de ``pagamento'' das saídas. Ocorre que:\\ 
\noindent (i) o fluxo de caixa de saída não é determinístico, depende de mortes, entradas em invalidez, aposentadorias, resgates, portabilidades e realocações de investimento, inclusive por mudanças de perfil feitas por participantes. Ainda que seja tentada uma gestão para imunizar essas saídas, sempre haverá o risco de transferência de riqueza de saída;\\
\noindent (ii) adicionalmente, saídas e realocações (que impliquem venda de títulos) podem ocorrer a qualquer momento ao longo do ano, enquanto que vencimentos de títulos ocorrem em datas específicas. Esse descasamento de prazo, por mais que se tente que seja pequeno, sempre levará a transferências de riqueza de saída, porque o valor na curva não será equivalente ao preço de mercado (apenas na data exata de vencimento);\\
\noindent (iii) ainda que seja adotada uma estratégia de imunização de alguma medida de risco das saídas (que são estocásticas), esse tipo de estratégia tem que ser revisitada frequentemente para se manter atualizada. Essas atualizações levam a realocações de ativos, ou seja, compras e vendas de títulos, que também causam transferências de riqueza caso o plano adote HTM.\\

Por mais que seja tentada a diminuição da transferência de riqueza de saída, ainda haverá a transferência de riqueza de entrada. Essa ocorre no momento da entrada de quaisquer recursos no plano pela diferença entre preços de mercado e valores na curva dos títulos públicos presentes na carteira quando da entrada ou realocação que leve à aquisição de títulos.

Planos CD entregam aos participantes todo o risco de retornos dos ativos. Não deveriam também impor transferências de riqueza (de entrada ou de saída) apenas por uma escolha de um determinado método para marcação de ativos e, consequentemente, cotização. Tal transferência pode trazer significativos prejuízos financeiros aos participantes advindos apenas por causa de um modelo de mensuração de títulos públicos federais sobre o qual o participante não tem qualquer responsabilidade por escolha (ao contrário do risco do perfil de seu investimento). Além disso, EFPCs não possuem capital próprio (formado por recursos não aportados por participantes) para arcar com um risco de prejuízo aos participantes vindo de uma escolha única de sua gestão.

\section{Considerações finais}

No contexto dos planos de previdência complementar nas modalidades de Contribuição Definida (CD) e Contribuição Variável (CV), na fase de diferimento, a metodologia utilizada para mensurar os ativos, especialmente os de renda fixa, e calcular o valor das cotas influencia diretamente se haverá ou não transferência de riqueza entre os participantes. A escolha entre a marcação a mercado (MTM – {\it mark to market}) e a marcação na curva (HTM – {\it hold to maturity}) é um dos pontos centrais com impactos significativos sobre a equidade nesses planos.

A regulamentação em vigor (Res. CNPC 61, de Dez/2024) passou a permitir que planos CD e CV, na fase de diferimento, de EFPCs utilizem a marcação na curva (HTM) de títulos públicos federais. Diante desse contexto, este artigo buscou quantificar as transferências de riqueza decorrentes dessa adoção, utilizando dados reais da estrutura a termo da taxa de juros de cupom de IPCA (ETTJ de cupom de IPCA) observadas entre Dez/2005 e Dez/2024. O uso destas taxas aproxima o estudo da realidade da variação, entre os anos, dos preços dos ativos, ou seja, aproxima o estudo da realidade das transferências de riqueza que ocorreriam nesse período, dadas as demais condições de cada cenário.

Neste artigo, uma metodologia foi desenvolvida para quantificar a transferência de riqueza para diferentes cenários de planos. Exemplos numéricos também foram construídos para mostrar como é mensurada a transferência quando da entrada de recursos, para aqueles que permanecem no plano ao longo do período de análise, e também quando da saída. 

Planos de previdência, inclusive CD e CV, na fase de diferimento, experimentam constantes entradas e saídas, que demandam frequentes realocações, novos investimentos e desinvestimentos. Os resultados indicam que sempre há transferência de riqueza quando a marcação na curva (HTM) é usada, ainda que seja de forma parcial. Afinal, sempre há entradas ou saídas em um plano de previdência. Essas transferências de riqueza podem ser significativas. Elas tendem a ser ainda maiores com alocação maior em títulos de longo prazo e com maiores movimentações no plano (contribuições, aportes, saídas por qualquer motivo e realocações devido a mudanças de perfis também). Participantes, em casos especiais, também podem arbitrar ganhos por transferência de riqueza às custas dos demais.

Como resumo e regra de bolso, podemos afirmar que quando a marcação na curva (HTM) é adotada:\\
\noindent (i) se as taxas de juros subirem, quem retira recursos do plano (ou leva o plano a vender títulos) tende a ter ganhos. Quem entra com recursos (ou aquisição de títulos) tende a ter perdas nesta parcela.\\
\noindent(ii) se as taxas de juros diminuírem, quem retira recursos tende a ter perdas. Quem entra com recursos tende a ter ganhos nesta parcela.\\
Como exposto no artigo, o objetivo do estudo não é expressar opinião sobre um ou outro modelo de mensuração de ativos, em específico, de títulos públicos. O estudo é baseado em desenvolvimento matemático para quantificar a transferência de riqueza que a marcação na curva (HTM) em planos CD, e CV na fase de diferimento, acarreta. Essas transferências de riqueza são fáticas. O volume de perdas para alguns participantes pode ser expressivo como visto nos resultados.

\bibliographystyle{apalike}
\bibliography{references}

\appendix
\section{Gráficos suplementares}

\subsection{Transferência de Riqueza na entrada}

\begin{figure}[ht!]
\caption{Superfície da perda máxima (\%) dos participantes que permanecem no plano até Dez/2024 devido às transferências de riqueza ocorridas com a marcação na curva (HTM) em relação à alocação em títulos de curto, médio e longo prazos (considerando taxa de entrada de $3,33\%$ e de saída de $0,10\%$). Número inicial de participantes = 10.000. Salário de contribuição = R\$15.000. Para pagamento de saídas, são vendidos os títulos de vencimentos mais curtos na carteira.}
\centering

\includegraphics[width=11cm]{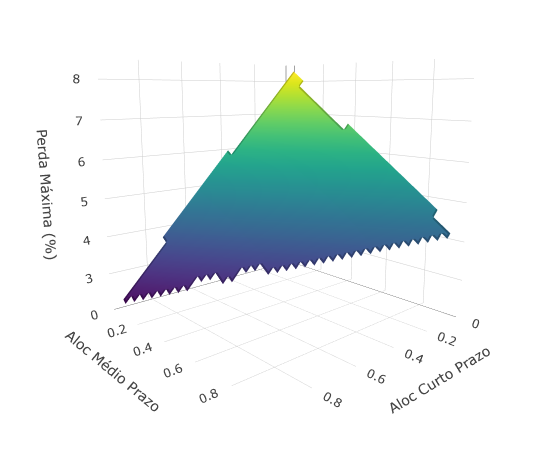}


\label{fig:fica_aloc}
\end{figure}




\begin{figure}[ht!]
\caption{Histogramas das Perdas (Ganhos) - valores (i) absolutos e (ii) \% - dos participantes que permanecem no plano até Dez/2024 devido às transferências de riqueza ocorridas com a marcação na curva (HTM). Número inicial de participantes = 10.000. Salário de contribuição = R\$15.000. Taxa de entrada = $3,33\%$. Taxa de saída = $0,10\%$. Alocação = $[25\%, 25\%, 50\%]$. Para pagamento de saídas, são vendidos os títulos de vencimentos mais curtos na carteira.}
\centering

\includegraphics[width=7cm]{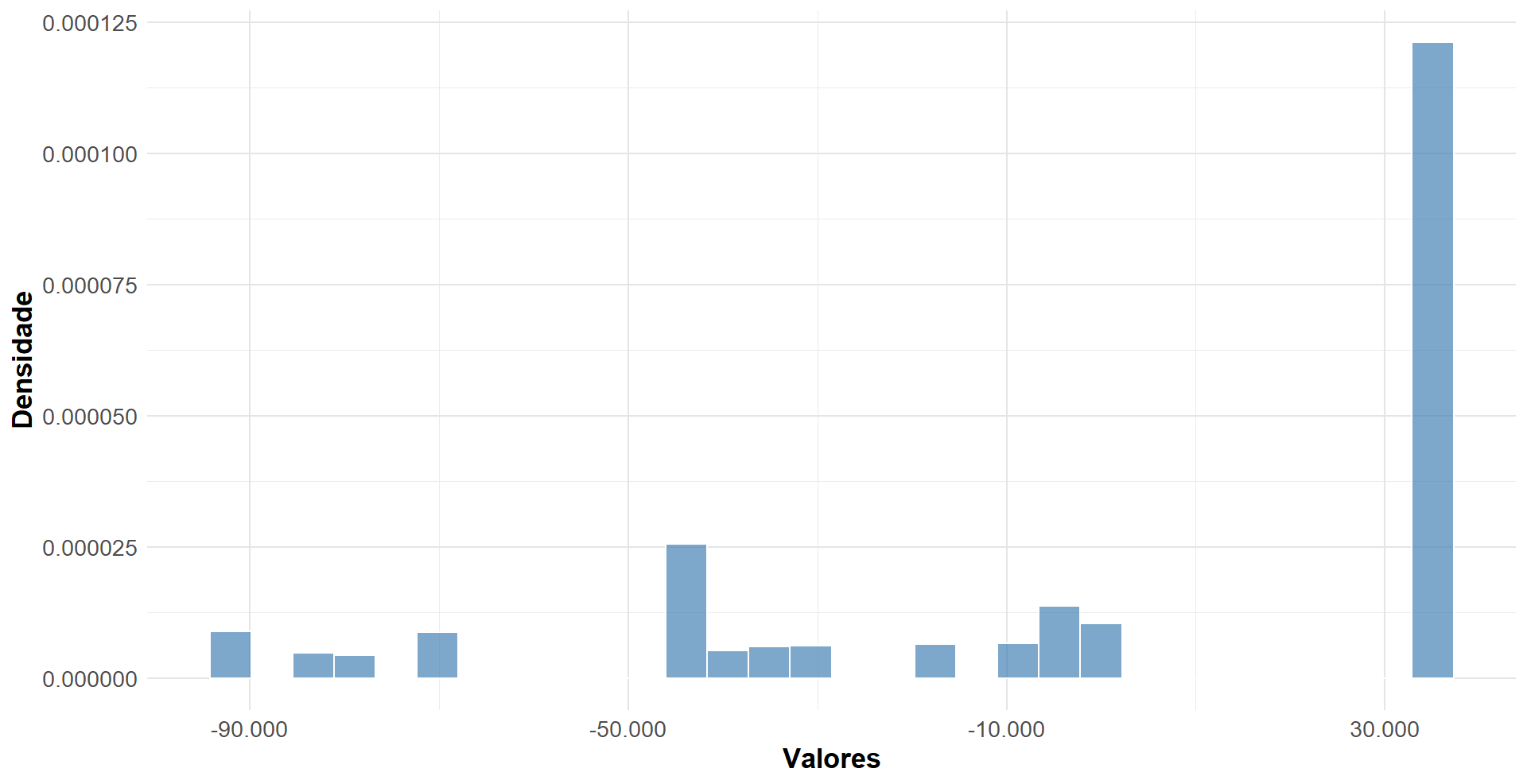}
\includegraphics[width=7cm]{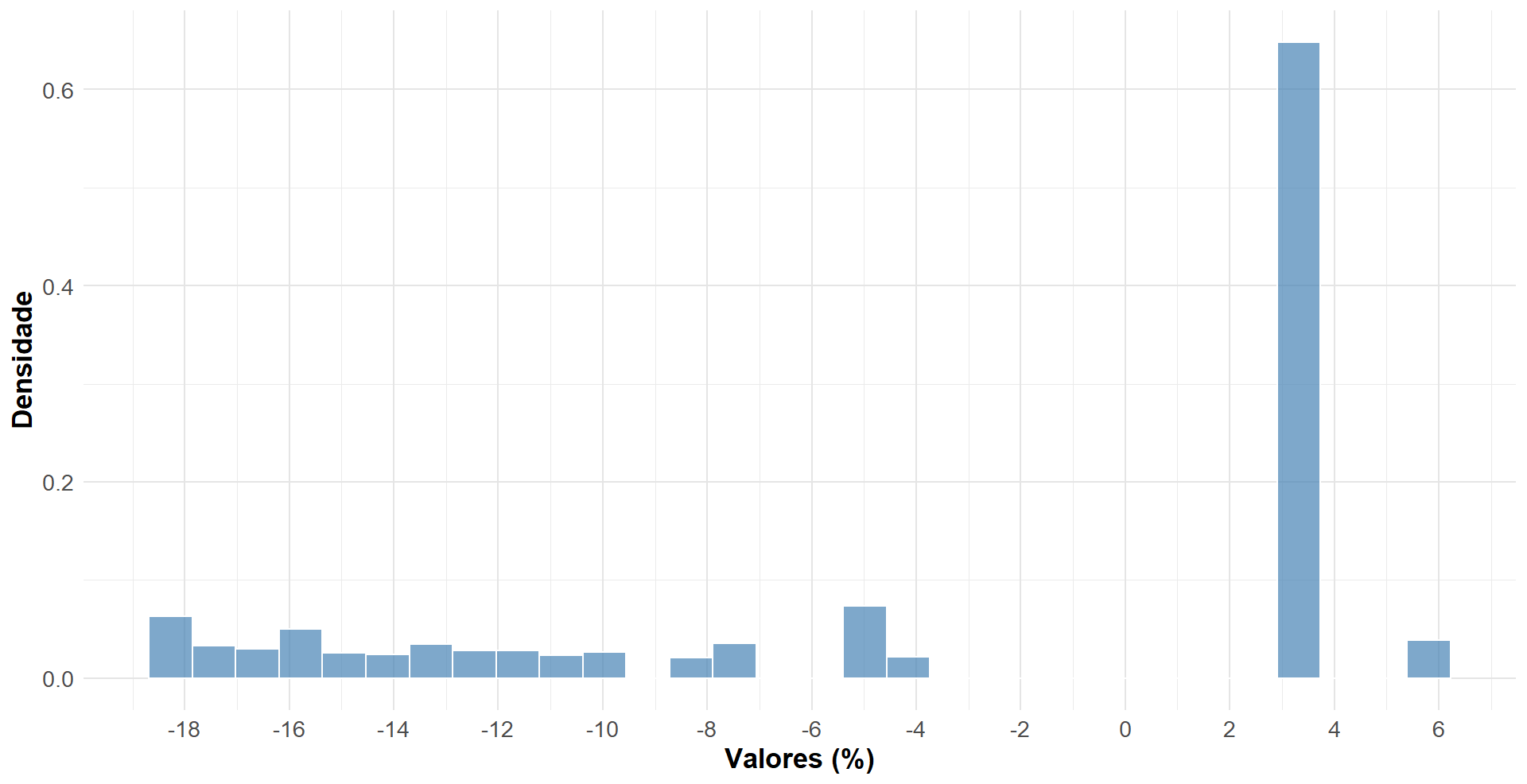}

\label{fig:fica_hist}
\end{figure}


\subsection{Transferência de Riqueza na saída}

\begin{figure}[ht!]
\caption{Superfície da perda máxima (\%) dos participantes que retiram recursos do plano devido às transferências de riqueza ocorridas com a marcação na curva (HTM) em relação à alocação em títulos de curto, médio e longo prazos (considerando taxa de entrada de $3,33\%$ e de saída de $0,10\%$). Número inicial de participantes = 10.000. Salário de contribuição = R\$15.000. Para pagamento de saídas, são vendidos os títulos de vencimentos mais curtos na carteira.}
\centering



\includegraphics[width=11cm]{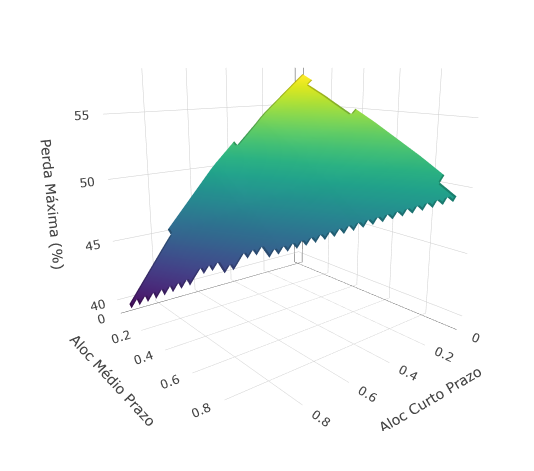}


\label{fig:saida_aloc}
\end{figure}


\begin{figure}[ht!]
\caption{Histogramas das Perdas (Ganhos) - valores (i) absolutos e (ii) \% - dos participantes que retiram recursos do plano devido às transferências de riqueza ocorridas com a marcação na curva (HTM). Número inicial de participantes = 10.000. Salário de contribuição = R\$15.000. Taxa de entrada = $3,33\%$. Taxa de saída = $0,10\%$. Alocação = $[25\%, 25\%, 50\%]$. Para pagamento de saídas, são vendidos os títulos de vencimentos mais curtos na carteira.}
\centering

\includegraphics[width=7cm]{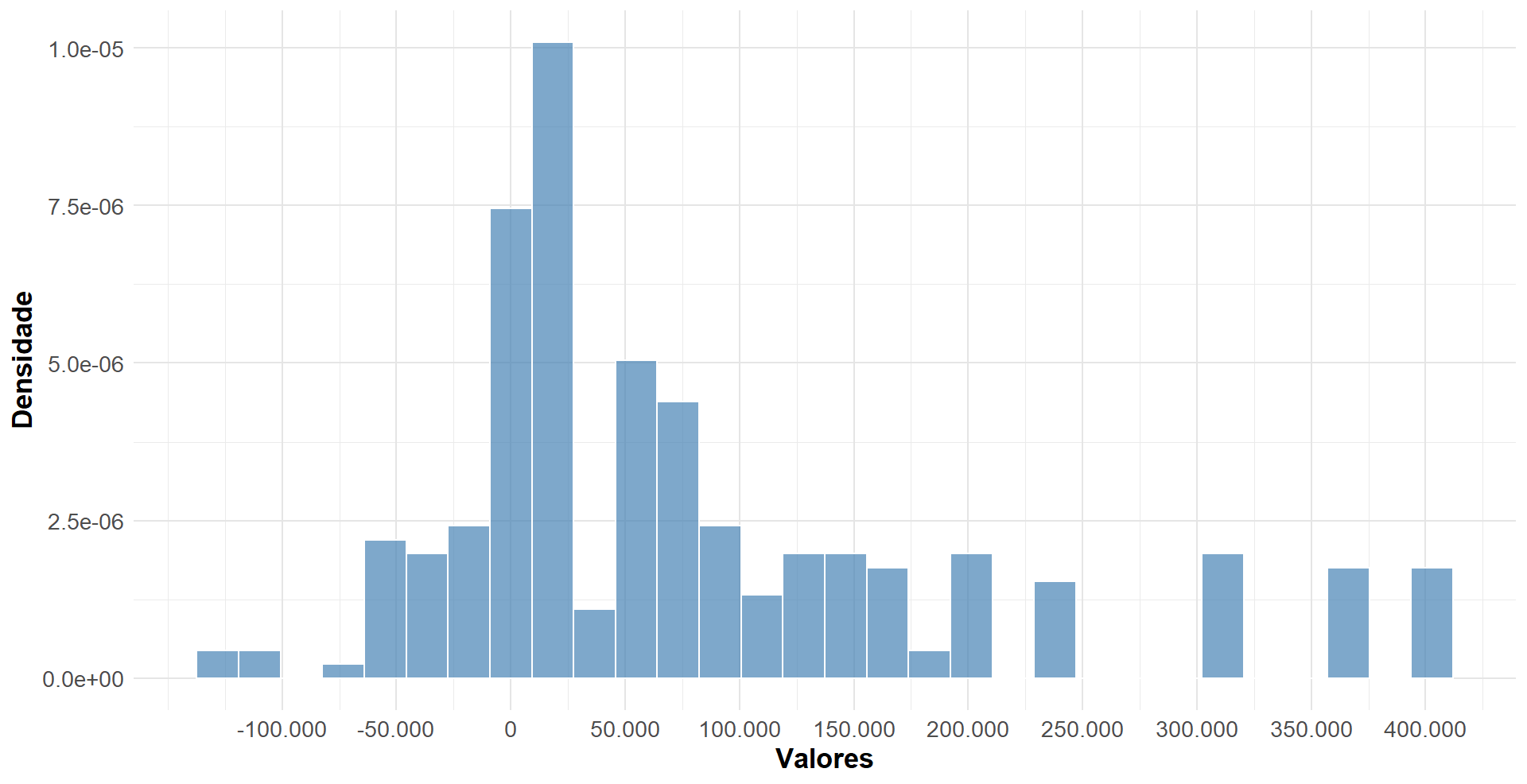}
\includegraphics[width=7cm]{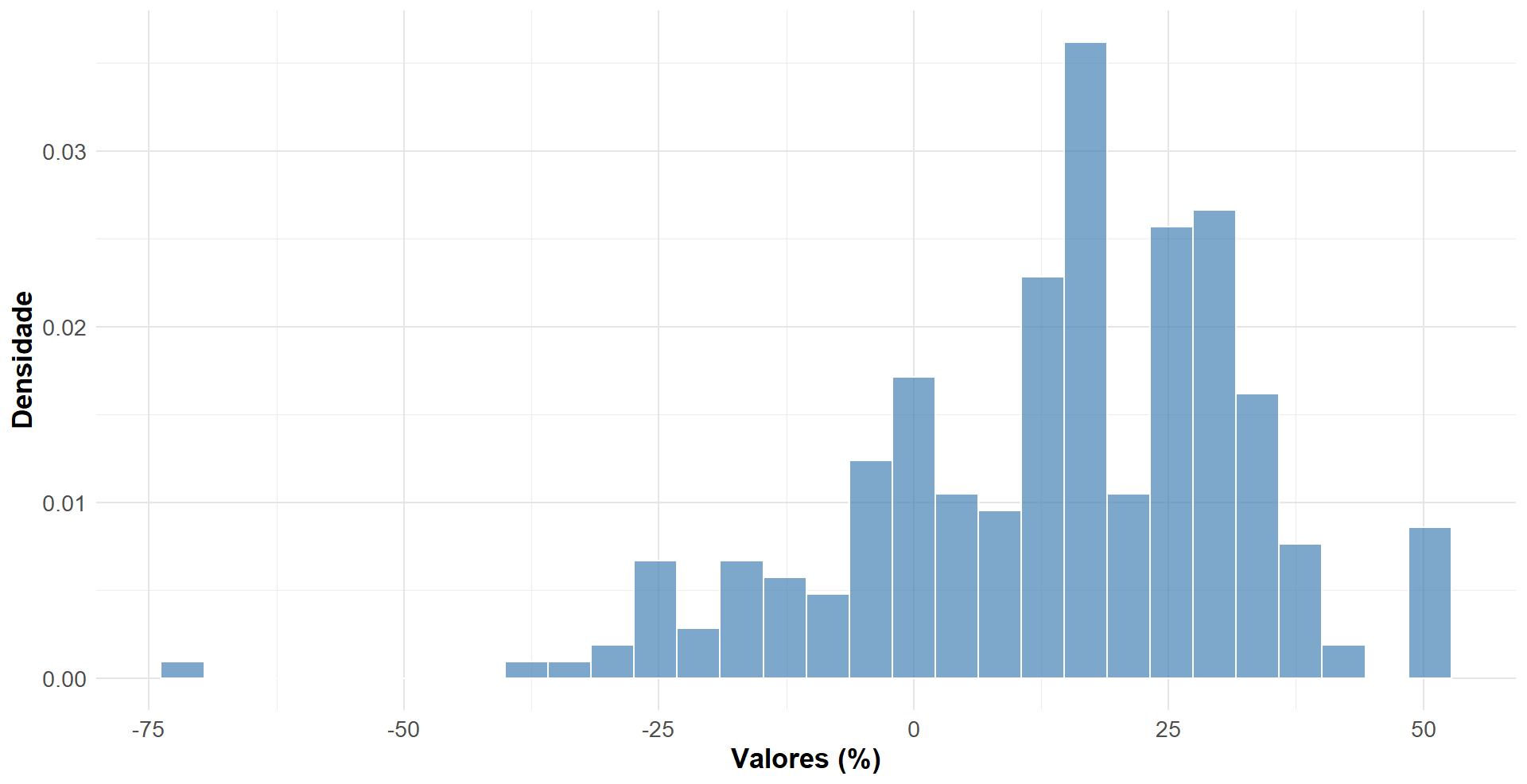}

\label{fig:saida_hist}
\end{figure}


\end{document}